\documentclass[a4paper,11pt]{article}
\usepackage{setspace}
\usepackage[left = 2cm, right=2cm, top=3cm, bottom=3cm]{geometry}
\usepackage[pdftex]{graphicx}
\usepackage{cite}
\usepackage{bm}
\usepackage{float}
\usepackage{amsmath,amssymb,latexsym}
\usepackage{color}
\usepackage{pdfpages}
\usepackage[T1]{fontenc}
\usepackage{wasysym}
\usepackage{siunitx}
\usepackage{setspace}
\usepackage{relsize}
\usepackage{tikz}
\usepackage{lineno, blindtext}
\usepackage{authblk}
\usepackage[symbol]{footmisc}
\usepackage{lineno}
\usepackage{chemformula}
\usepackage[pdfencoding=auto, psdextra]{hyperref}

\begin{document}
	\renewcommand{\figurename}{Fig.}
	\title{\color{blue}\textbf{Thermal History Asymmetry and Dissipation in Dense Colloidal Microgel Glasses}}
	\author[1]{Sonali Vasant Kawale}
	\affil[1]{\textit{Soft Condensed Matter Group, Raman Research Institute, C. V. Raman Avenue, Sadashivanagar, Bangalore 560 080, INDIA}}
	\author[2]{Yogesh M Joshi}
    \affil[2]{\textit{Department of Chemical Engineering, Indian Institute of Technology Kanpur, Kanpur, Uttar Pradesh, India, 208016. }}
	\author[1,*]{Ranjini Bandyopadhyay}
	
	\footnotetext[1]{Corresponding Author: Ranjini Bandyopadhyay; Email: ranjini@rri.res.in}
	\maketitle
	\begin{abstract}
{ Microstructurally arrested matter, from molecular glasses to soft glassy materials, can retain a memory of their thermal or mechanical (shear) histories. Their history-dependent and nonlinear microstructural recoveries have been studied within the Kovacs framework. Here, we applied the temperature ramps of varying magnitudes to dense colloidal suspensions of thermoresponsive, deformable and compressible microgel particles should serve as an effective strategy to probe the nonlinear path-dependent structural recovery of these systems. We synthesised Poly (N-isopropyl acrylamide) (PNIPAM) microgel particles using the free radical precipitation polymerisation method. Using oscillatory rheology, we studied the relaxations of the viscoelastic moduli of dense PNIPAM suspensions that were heated and cooled at various temperature ramp rates. Path-dependent structural recovery was quantified by studying the asymmetric approach of the suspension elastic modulus toward the target temperature during the heating and cooling temperature ramps. The loss modulus peaks, observed at the times of initiation and termination of the temperature ramps, were understood to arise from energy dissipation due to microgel rearrangement events. The heights of the peaks were found to be inversely correlated with the asymmetry in the elastic response. Our work highlights the important role of energy dissipation through microgel rearrangements in eliminating path-dependent asymmetries in the storage moduli of dense PNIPAM suspensions subjected to thermal shocks. By tuning the applied temperature ramp rate and particle packing density, therefore, asymmetric storage modulus relaxations in dense systems can be modulated \textit{via} adjustments of the accessible free volume.}

        \end{abstract}                   
	\noindent
	\definecolor{black}{rgb}{0.0, 0.0, 0.0}
	\definecolor{red(ryb)}{rgb}{1.0, 0.15, 0.07}
	\definecolor{darkred}{rgb}{0.55, 0.0, 0.0}
	\definecolor{blue(ryb)}{rgb}{0.01, 0.2, 1.0}
	\definecolor{darkcyan}{rgb}{0.0, 0.55, 0.55}
	\definecolor{navyblue}{rgb}{0.0, 0.0, 0.5}
	\definecolor{olivedrab(web)(olivedrab3)}{rgb}{0.42, 0.56, 0.14}
	\definecolor{darkraspberry}{rgb}{0.53, 0.15, 0.34}
	\definecolor{magenta}{rgb}{1.0, 0.0, 1.0}
	
	\newcommand{\blsquare}{\textcolor{black}{\small$\blacksquare$}}
	\newcommand{\hlsquare}{\textcolor{darkred}{\small$\square$}}
	\newcommand{\redtraingle}{\textcolor{magenta}{\small$\triangle$}}
	\newcommand{\oolive}{\textcolor{olivedrab(web)(olivedrab3)}{\large$\circ$}}
	\newcommand{\purpletraingle}{\textcolor{darkraspberry}{\small$\triangledown$}}
	
	\newcommand{\bltriangle}{\textcolor{black}{\small$\triangleup$}}
	\newcommand{\rcircle}{\textcolor{red(ryb)}{\large$\bullet$}}
	\newcommand{\rtraingle}{\textcolor{red(ryb)}{\small$\triangledown$}}
	\newcommand{\wine}{\textcolor{darkred}{\large$\bullet$}}
	\newcommand{\cyan}{\textcolor{darkcyan}{\large$\bullet$}}
	\newcommand{\owine}{\textcolor{darkred}{\large$\circ$}}
	\newcommand{\redcircle}{{\large$\circ$}}
	\newcommand{\ocyan}{\textcolor{darkcyan}{\large$\circ$}}
	\newcommand{\blue}{\textcolor{blue(ryb)}{\large$\bullet$}}
	\newcommand{\blbullet}{\textcolor{navyblue}{\large$\bullet$}}
	\newcommand{\olbullet}{\textcolor{olivedrab(web)(olivedrab3)}{\large$\bullet$}}
	\newcommand{\hollowblue}{\textcolor{blue(ryb)}{\large$\circ$}}
	\newcommand{\black}{\textcolor{black}{\large$\bullet$}}
	\newcommand{\hollowblack}{\textcolor{black}{\large$\circ$}}
	\newcommand{\bltria}{\textcolor{black}{\small$\triangle$}}
	\newcommand{\rtrai}{\textcolor{red(ryb)}{\large$\triangledown$}}	
	\section{Introduction:}
    
 {A system placed in a bath at a slightly different temperature will eventually attain thermal equilibrium in a quasi-static process. For a highly non-equilibrium system, relaxation to the equilibrium state is much more intricate, with strong, path-dependent asymmetries arising between the heating and cooling processes. An analysis of the responses of these systems during heating and cooling highlights their inherent asymmetries and underscores the necessity of investigating the fundamental mechanisms that drive variations in material behaviour. Such studies are crucial in advancing our knowledge in fields such as materials engineering and thermodynamics. 

Colloids are akin to scaled-up atoms and are used as model systems 
to mimic hard condensed matter. In this context, dense colloidal
suspensions have been widely used to model molecular glass
phenomenologies \cite{pusey1986phase, mattsson2009soft, di2011signatures, vlassopoulos2014tunable}. Some colloidal systems show a systematic and reversible thermoresponsive character and can undergo transitions between soft solid and liquid states when their temperatures are changed appropriately \cite{ basak2013encapsulation, negi2014viscoelasticity,suman2020universality, suman2021rheological, suman2022anomalous}. Among others, poly (N-isopropylacrylamide) (PNIPAM) is a classic thermoresponsive synthetic colloidal system constituted by deformable and compressible microgel particles. Aqueous PNIPAM suspensions show a reversible swelling-deswelling transition due to the absorption and expulsion of water at a volume phase transition temperature (VPTT) of approximately 34$^{\circ}$C \cite{dufresne2004preparation, romeo2010temperature, franco2021glass}. The soft and deformable nature of these particles allows us to control the volume fractions of their suspensions simply by tuning the medium temperature at a fixed microgel number density. The dynamic swelling characteristics of PNIPAM microgels are influenced by a range of external stimuli such as environmental temperature, pressure, humidity, medium pH, and ionic strength, etc., allowing their applications in fields such as drug delivery and biomaterial engineering \cite{romeo2010temperature, wu2003phase,lyon2012polymer, karg2008temperature, peppas2006hydrogels, misra2020influence,  vialetto2024effect, gury2025internal,  zhang2025unveiling}. The rheology of soft, thermoresponsive PNIPAM-based microgel suspensions has recently been comprehensively reviewed \cite{franco2025soft}. 

Using thermodynamic perturbation theory, light scattering measurements and spectroscopic analyses, Wu \textit{et al.} proposed a phase diagram in the temperature-microgel concentration plane \cite{wu2003phase}. 
In results that are reminiscent of those reported for molecular glasses, Mattsson \textit{et al.} \cite{mattsson2009soft} demonstrated that the fragilities of deformable microgels were determined by the elasticities of individual particles.  Romeo \textit{et al.} \cite{romeo2010temperature} demonstrated that concentrated PNIPAM suspensions exhibit interesting temperature-dependent rheology. As the solvent became increasingly unfavourable when the temperature was increased beyond the VPTT, the authors reported a transition from a colloidal glass to a colloidal gel phase and an intervening liquid-like phase at the VPTT. When the weight concentration of PNIPAM was varied in a PNIPAM-PAAc system while keeping the PAAc content fixed, three different rheological regimes, a shear-thinning fluid, an attractive glass characterised by a yield stress, and a jammed state, were identified \cite{franco2021glass}. Recently,  Misra \textit{et al.} \cite{misra2024effect} used rheological studies and scanning electron microscopy to demonstrate that the microstructural details and bulk rheology of dense suspensions of microgel particles were highly sensitive to changes in particle stiffness. 

The approach of a material towards its glass transition is known to show path dependence, and such structural recovery has been demonstrated to be ultra-slow. Landmark experiments developed by Kovacs to study non-equilibrium polymer glasses demonstrated that the time-dependent evolution of a glassy system is significantly impacted by its thermal history, and is characterised by asymmetric response, path dependence and memory effects
\cite{10.1007/BFb0050366, tool1946viscosity, struik1978physical, larson1999structure, bertin2003kovacs, yunker2009irreversible, joshi2014long, scalliet2019rejuvenation, edera2025mechanical}. 
These features have been investigated more extensively in later studies \cite{mckenna1999tau}. Experiments studying the properties of a glass during two temperature step changes of equal magnitude, an up-jump from T to T+$\Delta$T and a down-jump from T to T-$\Delta$T, demonstrated a lack of mirror symmetry in the volume recovery process \cite{10.1007/BFb0050366}. This asymmetry in structural recovery is a consequence of the highly nonlinear and path-dependent nature of the structural glass transition \cite{lira2021fundamental, mckenna201750th, kaushal2016analyzing, shukla2017boltzmann, agarwal2020signatures}. In polymer glasses, such asymmetry originates from the coupling between structural relaxation and dissipation, in such a way that the free-volume recovery and enthalpy relaxation in a system proceed along distinct, history-dependent pathways during heating and cooling cycles.

Structural recovery of a glassy thermoresponsive colloidal PNIPAM suspension was investigated by applying concentration jumps in diffusing wave spectroscopy measurements \cite{di2011signatures}. The approach of the system to the target temperature during up-down temperature jumps was noted to be asymmetric. Creep experiments, performed to investigate the physical aging and structural recovery of these systems, clearly demonstrated the similarities and disparities between colloidal and molecular glasses \cite{di2014dynamics,peng2016physical}. Despite self-similar evolution under isothermal conditions, a soft glassy clay dispersion exhibited asymmetric structural recovery upon temperature jumps due to restricted counterion mobility, resembling the Kovacs-like phenomenon in glasses \cite{Dhavale}. 

Strain amplitude jumps, applied to natural rubber samples reinforced with carbon black particles, displayed the same signatures of structural recovery as seen in glasses under temperature jumps \cite{robin2022glass}. Kovacs-like memory was observed in a combined rheology and X-ray photon correlation spectroscopy study of thermoresponsive nanocolloidal suspensions of attractive spheres undergoing gel formation and aging \cite {chen2023memory}. The results were verified by implementing a cluster model for gel formation. Chen \textit{et al.} modelled the effects of deformation protocols on the nonlinear mechanical response of polymer glasses by implementing a nonlinear Langevin equation (NLE) theory of segmental relaxation \cite{chen2010theory}. By altering the balance between aging and mechanical disordering \textit{via} application of different stresses and temperature profiles, the researchers were able to fully rejuvenate the system, thereby erasing the thermal history in the non-equilibrium steady state. In a recent work \cite{janzen2024rejuvenation}, researchers demonstrated that thermal cycling in active systems results in both rejuvenation and memory effects. However, for experiments with passive systems under the same conditions, memory effects and a complete absence of rejuvenation were reported.

In the present work, we hypothesised that the mechanical response of dense PNIPAM suspensions can be fine-tuned during heating and cooling processes by applying temperature ramps at different rates or by applying a fixed temperature ramp rate to suspensions of various concentrations. Deformable and compressible PNIPAM particles, synthesised with a polar crosslinker in a free radical precipitation polymerisation process, were observed to shrink abruptly due to rapid expulsion of water in a narrow temperature range just below the VPTT of 34$^{\circ}$C. We characterised the relaxation of the dense PNIPAM suspensions under temperature ramps applied at different rates by measuring their viscoelastic moduli, G$^\prime$ (elastic modulus) and G$^\prime$$^\prime$ (viscous modulus), in oscillatory rheological experiments. 

 For systems approaching a target temperature at slow ramp rates, we observed a strong asymmetry in the relaxation of the storage modulus during heating and cooling processes. We observed distinct peaks in the loss modulus data at the times of initiation and termination of both heating and cooling temperature ramps. The heights of the loss modulus peaks increased with ramp rate and, therefore, with faster acceleration of temperature. We interpreted the peaks as signatures of energy dissipation resulting from microgel rearrangements in response to the imposed thermal shocks. For experiments performed for a range of ramp rates and PNIPAM concentrations, we uncovered an inverse correlation between the asymmetry in the storage modulus response and the peak height of the loss modulus response recorded in the same experiment. The structural recovery of colloidal microgel glasses can therefore be accelerated in a controlled manner by controlling the rate of rearrangements of the constituent particles. We conclude that the slow glassy dynamics of dense suspensions of deformable and compressible colloidal particles can be fine-tuned by adjusting colloidal concentrations and applied temperature ramp rates in oscillatory rheological experiments.}

\section{Material and Methods}
\subsection{Synthesis of PNIPAM microgel particles}
  PNIPAM microgel particles were synthesised using a polar crosslinker \textit{via} the free radical precipitation polymerisation method \cite{mcphee1993poly}. A crosslinker concentration of 10$\%$ relative to the total NIPAM monomer mass was used for synthesis. All chemicals were procured from Sigma-Aldrich and used as received without any additional purification. In the polymerisation reaction, 7.0 g N-isopropylacrylamide (NIPAM) (99$\%$), 0.7 g of the cross-linker N,\textit{N$^\prime$}-methylenebis-
   acrylamide (MBA) (99.5$\%$), and 0.03 g SDS were dissolved in 470 ml of Milli-Q water (Millipore Corp.) in a three-neck round-bottom (RB) flask. This flask was equipped with a reflux condenser, a heating assembly with a magnetic stirrer (Heidolph), a platinum sensor, and a nitrogen gas (N$_2$) inlet and outlet. To remove dissolved oxygen, nitrogen gas was purged into the solution, and the mixture was stirred at 600 RPM for 30 minutes. Subsequently, the polymerisation reaction was initiated by adding 0.28 g of potassium persulfate (KPS) (99.9$\%$) dissolved in 30 ml of Milli-Q water at 70$^{\circ}$C into the reaction mixture. The mixture was stirred for four hours at 600 RPM to ensure thorough mixing of the components. After the reaction period, the suspension was allowed to cool to room temperature. Purification was performed by four successive centrifugations and re-dispersions at rotational speeds of 20,000 RPM for 60 min to remove SDS, the remaining monomers, oligomers, and impurities. The supernatant was removed after centrifugation, and the sample was dried by evaporating water. A fine powder was prepared by grinding the dried particles using a mortar and pestle. Glycerol (LR) was procured from S D Fine Chemicals Limited, Mumbai, and used as received.
\begin{figure}
    \centering
    \includegraphics[width=\linewidth]{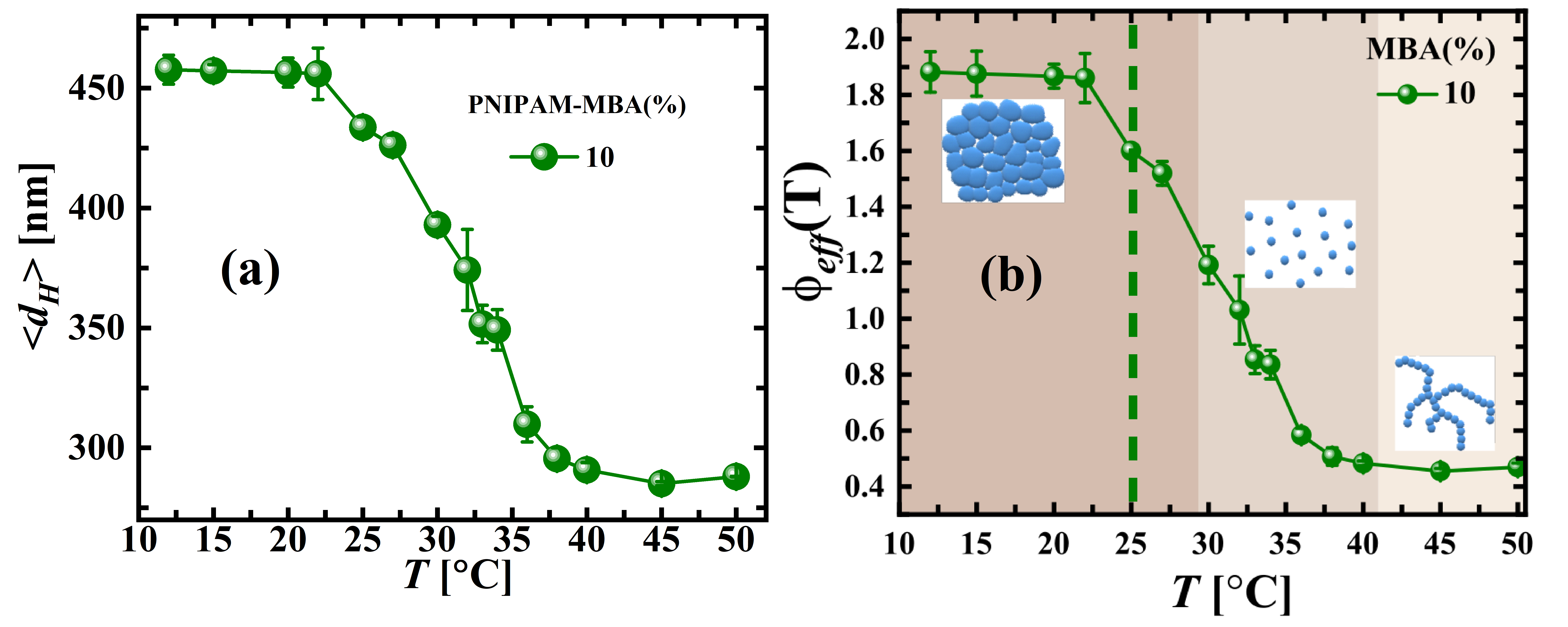}
    \caption{(a) Temperature-dependent average hydrodynamic diameters, $\langle d_H \rangle$, of PNIPAM particles in aqueous suspension. (b) Temperature-dependent effective volume fractions, $\phi_{eff}$(\textit{T}), of a dense aqueous suspension of PNIPAM particles, prepared at $\phi_{eff} =$ 1.6 at 25$^{\circ}$C (indicated by the vertical green dashed line). The insets from left to right display schematic illustrations of self-assembled microgel particles at temperatures below, near and above the VPTT. }
    \label{fig:schematicphieffective}
\end{figure}  
  
\subsection{Dynamic light scattering}
Dynamic light scattering (DLS) experiments were performed in a temperature range 12$^{\circ}$C - 55$^{\circ}$C, at intervals of 2$^{\circ}$C, to measure the average hydrodynamic diameters, $\langle d_H \rangle $, of the synthesised thermoresponsive PNIPAM particles. Details of the experimental setup, data acquisition, and analyses are provided in Section 1 of the supplementary material. A very dilute suspension (\(\phi \approx 10^{-3}\)) was prepared by adding PNIPAM microgel particles in Milli-Q water. 5 ml of each suspension was filled in a glass cuvette and loaded into the sample holder of the DLS setup. We computed the standard errors in our results by performing three repetitions of each experimental run. Fig.~\ref{fig:schematicphieffective}(a) shows the temperature-dependent $\langle d_H \rangle $ estimated from DLS measurements. As the temperature was increased towards the VPTT, we noted a rapid decrease in $\langle d_H \rangle$ due to expulsion of water. The temperature-dependent swelling ratio,  $\alpha_{T}$, of the PNIPAM particles was calculated using $\alpha_{T}$ = (\( d_{T^\circ\text{C}} \)/\( d_{50^\circ\text{C}} \)), where \( d_{T^\circ\text{C}} \) and \( d_{50^\circ\text{C}} \) are the average hydrodynamic diameters at temperatures \textit{T} and 50$^{\circ}$C, estimated using DLS and plotted in Fig. S1 of the supplementary material. The maximum swelling ratio, $\alpha$, is calculated using the formula $\alpha$ = (\( d_{20^\circ\text{C}} \)/\( d_{50^\circ\text{C}} \)), and is equal to 1.58 for the PNIPAM microgel particles used here.  

\subsection{Preparation of dense aqueous PNIPAM microgel suspensions}
Dense suspensions of PNIPAM microgel particles were prepared by adding PNIPAM powder to Milli-Q water. The samples were stirred continuously for 24 hours, sonicated for 15 minutes, and stored in a refrigerator maintained at a temperature of 4$^{\circ}$C for further use. Since PNIPAM particles are deformable and compressible, the volume fraction, $\phi$, is not an appropriate parameter to characterise their aqueous suspensions. To quantify particle packing in a dense PNIPAM suspension, a modified parameter, the effective suspension volume fraction $\phi_{eff}$, was estimated as described in Section 2 of the supplementary material. The concentrations of the ingredients were chosen to ensure an effective volume fraction $\phi_{eff}$=1.6 at $T=25^\circ \text{C}$. For the preparation of suspensions with different effective volume fractions, the ingredients were adjusted appropriately.  
We estimated temperature-dependent $\phi_{eff}(T)$ values \cite{carrier2009nonlinear} using the following relation:

\begin{equation}
{\phi_{eff}(T) = \phi_{eff}(25^\circ \text{C}) \times \left( \frac{d_h(T)}{d_h(25^\circ \text{C})} \right)^3, }
\end{equation}
where $\phi_{eff}(T)$ and $\phi_{eff}(25^\circ \text{C})$ are effective volume fractions, while $d_h(T)$ and $d_h(25^\circ \text{C})$ are average hydrodynamic diameters at temperatures $T$ and $25^\circ \text{C}$, respectively. We see from Fig.~\ref{fig:schematicphieffective}(b) that $\phi_{eff}$ is maximum below the VPTT, indicating a dense packing configuration of fully swollen microgel particles. The rapid decrease in $\phi_{eff}$ as VPTT is approached indicates shrinkage of the particles due to expulsion of water and subsequent particle unjamming. At temperatures above the VPTT, where inter-particle hydrophobic interactions dominate \cite{romeo2010temperature}, the collapsed particles attract to form gels. The distinct morphologies of self-assembled PNIPAM particles at different temperatures are shown schematically in the inset of Fig.~\ref{fig:schematicphieffective}(b).

\begin{figure}
    \centering
    \includegraphics[width=0.8\linewidth,height=0.8\textheight,keepaspectratio]{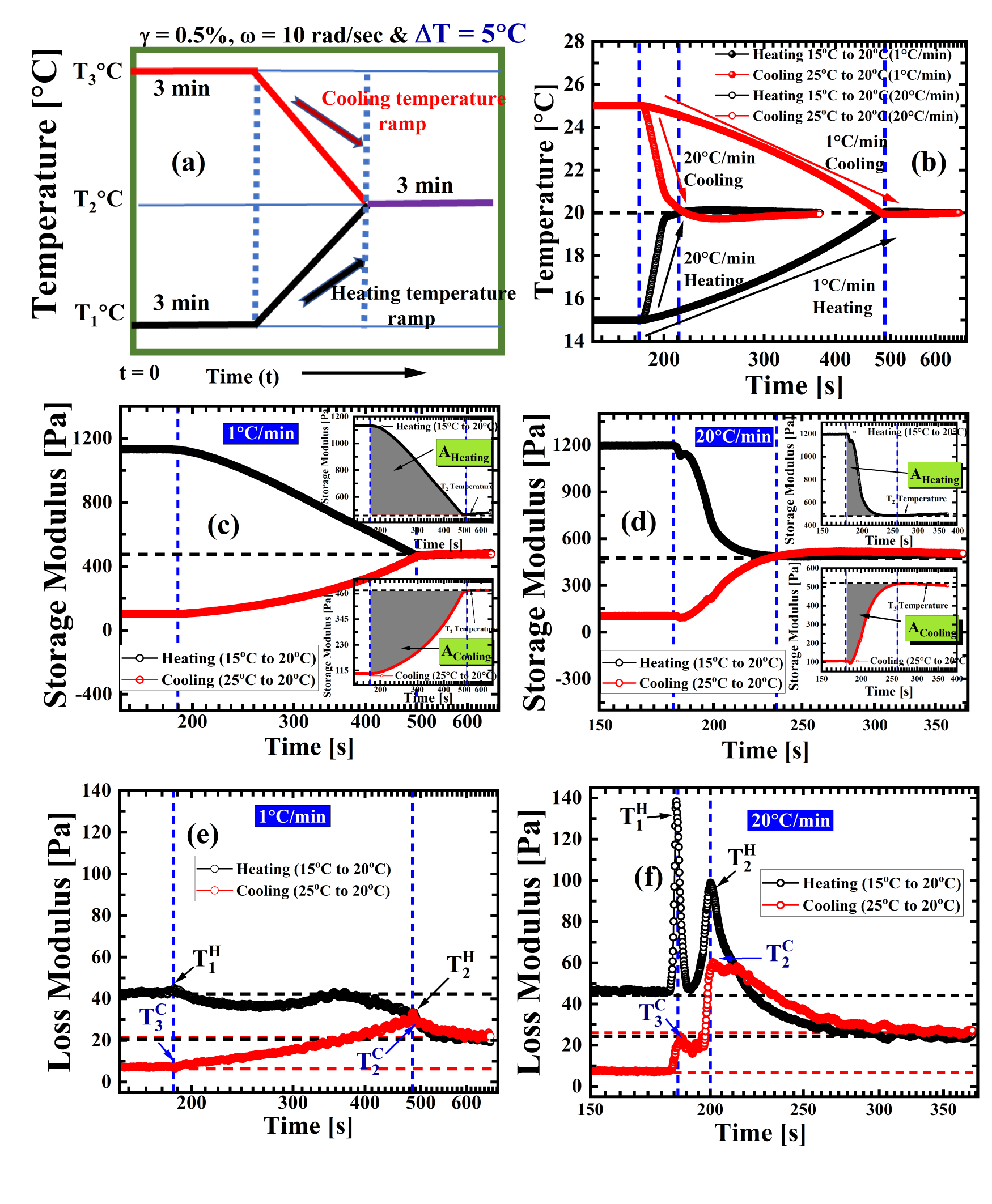}
    \caption{(a) Schematic of the experimental protocol used for heating and cooling ramp experiments. (b) displays temperature profiles applied during representative heating (15$^{\circ}$C to 20$^{\circ}$C) and cooling (25$^{\circ}$C to 20$^{\circ}$C) ramp experiments at ramp rates of 1$^{\circ}$C/min and 20$^{\circ}$C/min. (c)  and (d) respectively show the time-dependent responses of the storage modulus, G$^\prime$, of an aqueous PNIPAM suspension of effective volume fraction, $\phi_{eff}$ =1.6 at 25$^{\circ}$C, during heating and cooling temperature ramps applied at rates 1$^{\circ}$C/min and 20$^{\circ}$C/min. Insets in figures (c) and (d) illustrate the areas enclosed by the storage modulus-time plots (shaded in grey) during the heating and cooling ramp experiments, $A_{Heating}$ and $A_{Cooling}$, respectively. 
    (e) and (f) respectively show the loss modulus, G$^\prime$$^\prime$, responses measured simultaneously in the heating and cooling temperature ramp experiments. T$^H_1$ and T$^H_2$ refer to the loss modulus peaks around the temperatures $T_1$ and $T_2$ during the heating experiment, while T$^C_3$ and T$^C_2$ indicate the peaks around the temperatures $T_3$ and $T_2$ during the cooling experiment. In all the experiments, the temperature gradient, $\Delta$T, was kept fixed at 5$^\circ\text{C}$. }
    \label{fig:schematicProtocol}
\end{figure}

\subsection{Rheology}
Rheological measurements \cite{macosko1994rheology} were performed using a stress-controlled Anton Paar MCR 702 rheometer. A double gap geometry with a gap of 1.886 mm, an effective length of 40 mm, and requiring a sample volume of 3.8 ml was used to measure flow curves of dilute suspensions of PNIPAM particles. Flow curves were analysed to evaluate the effective volume fractions of the PNIPAM suspensions used in this work. Oscillatory rheological experiments were performed using a cone and plate geometry with a cone radius \(r_C\) = 12.491 mm, a cone angle of 0.979$^{\circ}$, measuring gap d = 0.048 mm, and requiring a sample volume of 0.07 ml. The cone and plate geometry was selected because of the relatively high viscosities of the samples studied here and the ability of the geometry to maintain a uniform strain rate throughout each sample during measurement. A Peltier unit was used to control sample temperature between 0$^{\circ}$C - 180$^{\circ}$C, while a water circulation system (Viscotherm VT2), capable of controlling temperature between 5$^{\circ}$C - 80$^{\circ}$C, was used with the double gap geometry. A solvent trap of silicone oil with a viscosity of 5 cSt was used to minimise water evaporation. We investigated the viscoelastic properties of dense PNIPAM suspensions under temperature ramps applied at different rates in oscillatory rheological experiments. To minimise the effects of sample loading and aging, we implemented a rejuvenation protocol prior to each measurement.

Fig.~\ref{fig:schematicProtocol}(a) illustrates the experimental protocol adopted by us for our rheological experiments. To ensure that the sample was initially in a liquid state, it was first heated to 35$^\circ\text{C}$. The sample was then cooled to a temperature of $T_1$$^\circ\text{C}$, where it was equilibrated for at least 5 minutes. After acquiring data at $T_{1}^{\circ}$C for 3 minutes, a temperature ramp was applied to heat the sample to another predetermined target temperature $ T_2$$^\circ\text {C}$. We designated this protocol, indicated by the black arrow in Fig.~\ref{fig:schematicProtocol}(a), as the heating ramp experiment. Next, the sample, now at $ T_2$$^\circ\text {C}$, was heated to 35$^\circ\text{C}$, and subsequently cooled to another predetermined temperature $T_3$$^\circ\text{C}$ ($T_3>T_2>T_1$). Data was acquired at $T_{3}^{\circ}$C for 3 minutes. We then implemented a cooling ramp experiment (indicated by the red arrow in Fig.~\ref{fig:schematicProtocol}(a)) wherein the sample was cooled to the target temperature $T_2$$^\circ\text{C}$ at the same temperature ramp rate as applied in the heating ramp experiment. 
The temperature difference $\Delta$T$^{\circ}$C (= $T_2$-$T_1$ and $T_3$-$T_2$)  was therefore maintained at a fixed value of 5$^{\circ}$C during the heating and cooling ramps. After completion of ramp applications in each experiment, the data was recorded for an additional 3 minutes.

The oscillatory rheological experiments reported here were performed at a constant applied shear deformation, $\gamma$ $=$ 0.5$\%$, and an angular frequency of 10 rad/sec. We implemented heating and cooling experiments (i) at different ramp rates using dense PNIPAM suspensions prepared with $\phi_{eff}$ = 1.6 at 25$^{\circ}$C, and (ii) by subjecting PNIPAM suspensions, prepared at different effective volume fractions at 25$^{\circ}$C, to a fixed temperature ramp rate. The elastic moduli, G$^\prime$, and the loss moduli, G$^\prime$$^\prime$, were monitored throughout.

\section{Results and Discussion}
\subsection{Viscoelastic moduli measured at different temperature ramp rates}
{In the first set of rheological experiments, we used dense aqueous suspensions of PNIPAM microgel particles, prepared at a fixed effective volume fraction, $\phi_{eff}$ =1.6 at 25$^{\circ}$C. Fig.~\ref{fig:schematicProtocol}(b) displays the temperature profiles for two representative heating and cooling experiments ($T_1$ $=$ 15$^{\circ}$C, $T_2$ $=$ 20$^{\circ}$C and $T_3$ $=$ 25$^{\circ}$C) at temperature ramp rates 1$^{\circ}$C/min and 20$^{\circ}$C/min. The two vertical blue dashed lines in each plot in Fig.~\ref{fig:schematicProtocol} indicate the times of initiation and termination of the applied temperature ramp. Figs.~\ref{fig:schematicProtocol}(c,d) show the relaxations of the storage modulus, G$^{\prime}$, during these heating and cooling experiments. We note a lack of mirror symmetry \cite{mckenna201750th} in G$^{\prime}$ as the system approaches $T_2$$^{\circ}$C under heating and cooling ramps. The time-dependent relaxations of the loss modulus, G$^{\prime\prime}$, measured simultaneously with G$^{\prime}$, are characterised by the appearance of two peaks centred around the initiation and termination points of the temperature ramps, as seen in Figs.~\ref{fig:schematicProtocol}(e,f). While G$^{\prime\prime}$ peaks are very broad and weak at 1$^{\circ}$C/min, they are stronger and sharper at 20$^{\circ}$C/min. For the heating ramp between $T_1$ and $T_2$, we designate the loss modulus peaks at the initiation and termination points of the ramp as T$^H_1$ and T$^H_2$, respectively. Similarly, for the cooling experiment between $T_3$ and $T_2$, the loss modulus peaks at the initiation and termination of the ramp are respectively labelled T$^C_3$ and T$^C_2$.}

 \begin{figure}]
    \centering
    \includegraphics[width=\linewidth]{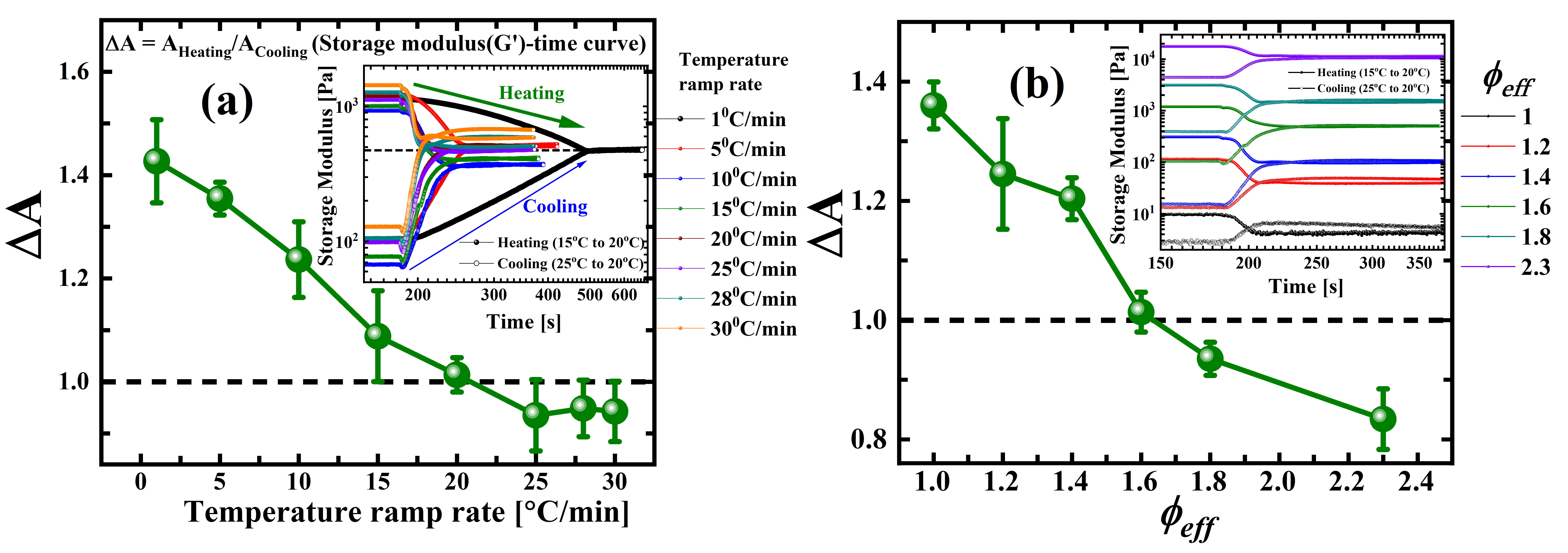}
    \caption{(a) Asymmetry of approach, $\Delta$A, of the storage modulus, G$^\prime$, towards $T_{2}^{\circ}$C. The data was acquired for an aqueous PNIPAM suspension of $\phi_{eff}$ =1.6 at 25$^{\circ}$C, at different heating and cooling ramp rates. (b) Asymmetry of approach of G$^\prime$ towards $T_{2}^{\circ}$C for heating and cooling experiments, applied at a fixed ramp rate of 20${^{\circ}}$C/min, in suspensions prepared at different effective volume fractions, $\phi_{eff}$, at 25$^{\circ}$C. The dashed black lines in both plots represent $\Delta$A = 1. The error bars denote standard deviations estimated from three independent measurements. The insets in (a) and (b) display raw storage moduli data for different temperature ramp rates and $\phi_{eff}$, respectively.} 
    \label{fig:schematicSMdifferentrates}
\end{figure}

\subsection{Asymmetry of approach of G$^\prime$ towards $T_2$}
To analyse the asymmetry of approach of the storage modulus, G$^{\prime}$, towards the target temperature, $T_2$, during heating and cooling temperature ramp experiments, we estimated an asymmetry parameter, $\Delta$A. We defined $\Delta$A as the ratio of the areas enclosed by the storage modulus versus time (G${^\prime}$ \textit{vs.} t) curves during heating and cooling ramps \cite{wang2015asymmetric, viollaz2003structural, ibanez2024heating} applied at the same rate. Therefore, $\Delta$A = $\frac{A_{Heating}}{A_{Cooling}}$, where the areas $A_{Heating}$ and $A_{Cooling}$ are highlighted in the insets of Figs.~\ref{fig:schematicProtocol}(c,d). This definition of asymmetry ensures that a larger deviation of $\Delta$A from 1 indicates greater asymmetry in the G$^\prime$ response. The inset of Fig.~\ref{fig:schematicSMdifferentrates}(a) shows the time-dependent G$^\prime$ data, measured during both heating and cooling experiments, at several ramp rates between 1$^{\circ}$C/min and 30$^{\circ}$C/min. Fig.~\ref{fig:schematicSMdifferentrates}(a), which displays the asymmetries, $\Delta$A, calculated for all applied ramp rates, clearly highlights two major features of the approach of G$^\prime$ towards $T_{2}$: (i) asymmetry is highly sensitive to the imposed temperature ramp rate for a fixed value of $\Delta$T, and (ii) asymmetry disappears at temperature ramp rates $\geq$ 15$^{\circ}$C/min, as indicated by $\Delta$A = 1 (black dashed line in Fig.~\ref{fig:schematicSMdifferentrates}(a)). 

Higher asymmetries at lower temperature ramp rates, as seen in Fig.~\ref{fig:schematicSMdifferentrates}(a), indicate strongly path-dependent mechanical responses \cite{mckenna201750th}. Such path dependence is a consequence of the distinct energy landscapes that the system navigates during the heating and cooling processes.
As the particles swell during the cooling ramp by absorbing water from the medium, the increasingly jammed configurations of the swollen microgels result in a highly rugged potential energy landscape. Such landscapes are very difficult to navigate, with pathways to the target temperature, $T_2$, becoming highly intricate. Deswelling of the microgel particles during the heating ramp, in contrast, results in particle shrinkage due to expulsion of water, thereby enhancing the availability of free volume and facilitating relaxation of the storage modulus. Simultaneously, the deformability and compressibility of the microgel particles expedite local dynamics.  At low ramp rates, therefore, the approach of G$^{\prime}$ towards $T_2$ proceeds along very distinct paths during heating and cooling temperature ramps, which leads to the observed asymmetry. Figs. S3(b-d) of the supplementary material shows that the maximum accelerations and decelerations of temperature, $\frac{d^{2}T}{dt^{2}}$, occur at the initiation and termination of the temperature ramp and increase with increasing ramp rate. Comparison between Figs.~\ref{fig:schematicSMdifferentrates}(a) and S3(d) reveals an inverse correlation between the asymmetry parameter, $\Delta$A, and the acceleration and deceleration of temperature, $\frac{d^{2}T}{dt^{2}}$, at the initiation and termination of the temperature ramp. 

We next performed temperature ramp experiments at 20$^{\circ}$C/min and $\Delta$T = 5$^{\circ}$C  with suspensions prepared at different effective volume fractions, $\phi_{eff}$, at 25$^{\circ}$C. Fig.~\ref {fig:schematicSMdifferentrates}(b) shows the approach of the storage modulus, G$^{\prime}$, towards $T_2$ for suspensions prepared at effective volume fractions varying between $\phi_{eff}$ = 1-2.3 at 25$^{\circ}$C. The inset in Fig.~\ref{fig:schematicSMdifferentrates}(b) displays the raw G$^{\prime}$ data during both heating and cooling processes. The asymmetry parameter, $\Delta$A, calculated as described earlier, decreases with increase in $\phi_{eff}$. As previously discussed, free volume is created rapidly due to microgel particle deswelling when PNIPAM suspension temperature is increased. We note that the larger availability of free volume at lower $\phi_{eff}$ is reminiscent of the effects of suspension heating. Increasing $\phi_{eff}$, in contrast, is akin to cooling the suspension, with the constituent PNIPAM particles becoming increasingly constrained in both scenarios. This relationship highlights the dynamic interplay between temperature and particle confinement in thermoresponsive PNIPAM suspensions. To conclude, swelling and deswelling processes occur along different pathways in these systems and allow navigation through distinct routes in the potential energy landscape \cite{stillinger2002energy, hornig2012static}. This results in the observed asymmetries in G${^\prime}$ when temperature ramp rate and particle concentration are appropriately tuned. 

Increasing $\phi_{eff}$ beyond 1.6 drives the system towards kinetic arrest. This inhibits microgel dynamics due to the non-availability of free volume and restricts the relaxation of the system towards T$_2$. We note that suspensions at the highest effective volume fractions are expected to be in a deeply supercooled or glassy state. The observed loss of asymmetry in these samples, as shown in Fig.~\ref{fig:schematicSMdifferentrates}(b), is a manifestation of their non-ergodic nature.

\begin{figure}
    \centering
    \includegraphics[width=0.7\linewidth]{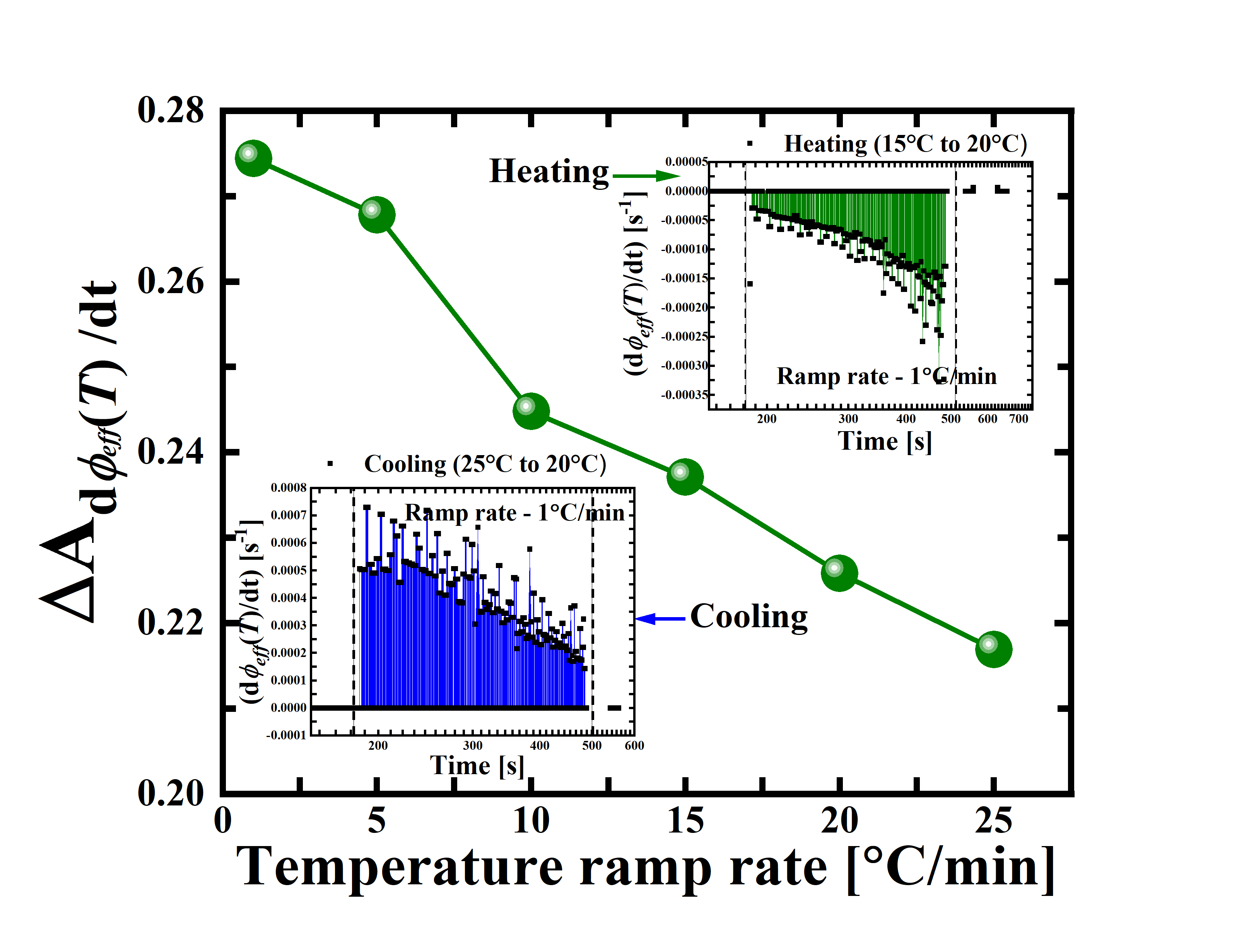}
    \caption{ Insets show the temporal changes in rate of change of effective volume fraction with time, $\frac{d\phi_{\textit{eff}}{(T)}}{dt}$, during heating (15$^{\circ}$C to 20$^{\circ}$C) and cooling (25$^{\circ}$C to 20$^{\circ}$C) ramps at a representative ramp rate of  1$^{\circ}$C/min. The green and blue shaded regions indicate the areas enclosed by the $\frac{d\phi_{\textit{eff}}{(T)}}{dt}$-time curves during heating and cooling processes, respectively. The main figure displays the asymmetric approach of $\frac{d\phi_{\textit{eff}}{(T)}}{dt}$, $\Delta A_{\frac{d\phi_{\textit{eff}}{(T)}}{dt}}$, towards the target temperature under heating and cooling temperature ramps, applied at rates between 1$^{\circ}$C/min and 25$^{\circ}$C/min. }
    \label{fig:schematicphieffectiveAsymmetry}
\end{figure} 
\subsection{Asymmetry in $\frac{d\phi_{\textit{eff}}{(T)}}{dt}$ during heating and cooling ramps}
To connect the observed asymmetry in G$^\prime$ to the free volume available to the microgel particles, we calculated the rate of change of the temperature-dependent effective volume fraction with time, $\frac{d\phi_{\textit{eff}}{(T)}}{dt}$, during heating and cooling ramps applied to a sample prepared at 25$^{\circ}$C with $\phi_{eff}$= 1.6. Detailed discussion of the analyses is provided in Section 3 of the supplementary material. With an increase in applied temperature ramp rates, we noted a decrease in the asymmetry of approach of $\frac{d\phi_{\textit{eff}}{(T)}}{dt}$ towards the target temperature $T_2$ during heating and cooling ramps. Raw data supporting this analysis can be found in Fig. S4 of the supplementary material. The corresponding asymmetry parameter, $\Delta A_{\frac{d\phi_{\textit{eff}}{(T)}}{dt}}$, plotted in Fig.~\ref{fig:schematicphieffectiveAsymmetry}, was estimated by taking the ratio of the areas enclosed by the $\frac{d\phi_{\textit{eff}}{(T)}}{dt}$-time curves during heating and cooling processes at different applied temperature ramp rates. The insets of Fig.~\ref{fig:schematicphieffectiveAsymmetry} highlight these areas of interest (green for the heating ramp and blue for the cooling ramp; data acquired at applied ramp rate 1$^{\circ}$C/min).  

We note that a reduction in $\phi_{eff}{(T)}$ indicates an increase in the free volume available for particle dynamics. Stronger asymmetry in $\frac{d\phi_{\textit{eff}}{(T)}}{dt}$ at a slower ramp rate implies that free volume is created during the heating ramp at a rate that is considerably faster than that for trapping/ freezing of free volume during the cooling ramp experiment. The asymmetric behaviour seen in Fig.~\ref{fig:schematicphieffectiveAsymmetry} therefore highlights the different rates at which free volume is created or trapped during heating and cooling ramps \cite{mckenna1999tau, peng2016physical}. These findings substantiate our claim that the relaxation of the suspension storage moduli towards temperature $T_2$ at different applied ramp rates can be correlated with the rates of change of free volume accessible to the microgel particles.

\begin{figure}
    \centering
    \includegraphics[width=\linewidth]{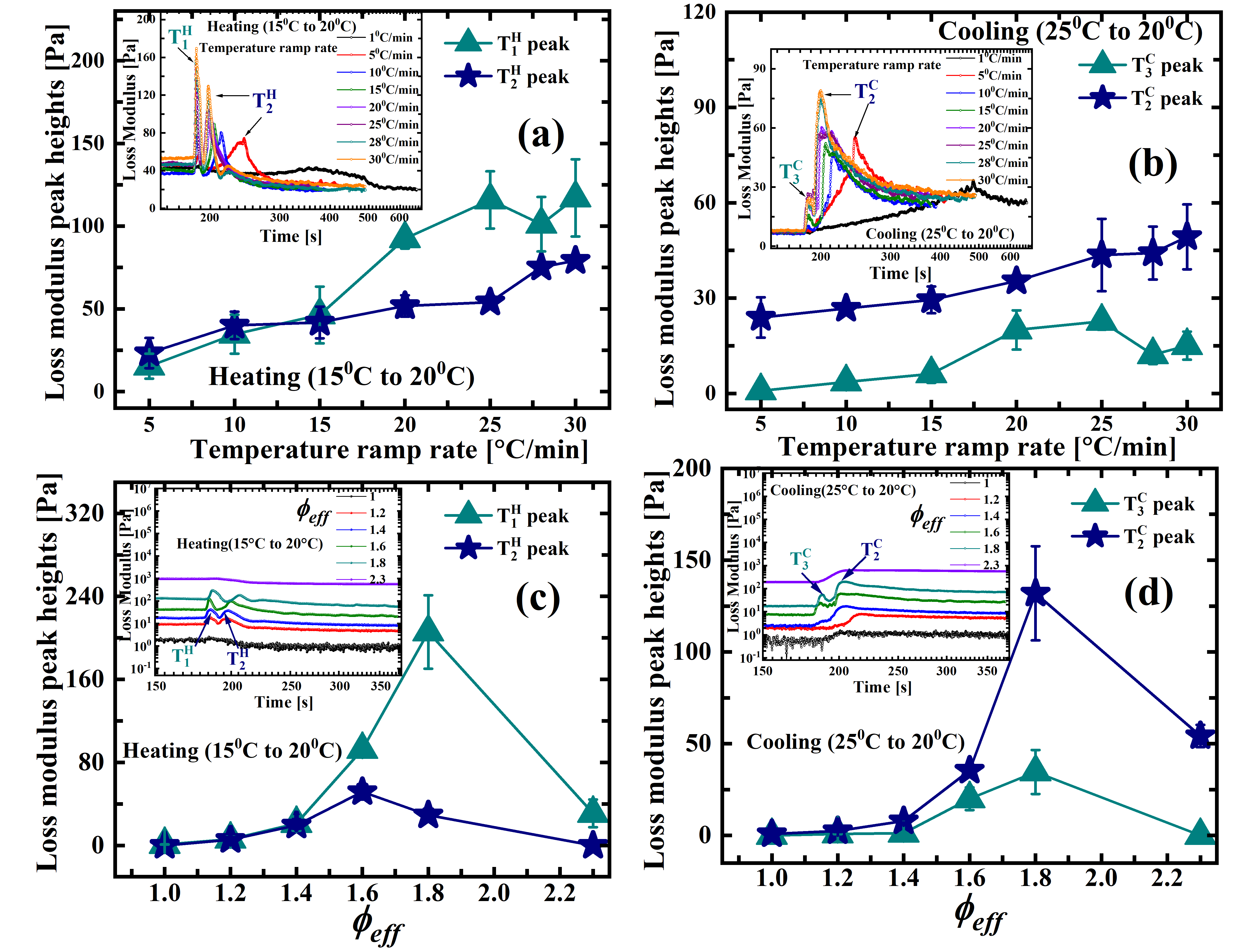}
     \caption{(a) Loss modulus, G$^{\prime\prime}$, peak heights when a PNIPAM suspension of effective volume fraction 1.6 at 25$^\circ$C was heated from 15$^\circ$C to 20$^\circ$C at different temperature ramp rates. The inset shows the raw loss modulus data during heating ramp experiments. (b) G$^{\prime\prime}$ peak heights when the suspension was cooled from 25$^\circ$C to 20$^\circ$C at the same applied ramp rates. The inset shows the corresponding raw loss modulus data. Loss modulus, G$^{\prime\prime}$, peak heights for different effective PNIPAM volume fractions (c) during heating and (d) cooling temperature ramps. Experiments in (c) and (d) were performed at a constant temperature ramp rate of 20$^\circ$C/min. The insets in (c) and (d) show raw loss modulus data during heating and cooling, respectively. T$^H_1$ and T$^H_2$ in (a) and (c) respectively represent the observed loss modulus peak heights at the blue{initiation} and termination of the heating ramp, while T$^C_3$ and T$^C_2$ in (b) and (d) respectively represent the loss modulus peak heights at the initiation and termination of the cooling ramp. The error bars were calculated from three independent measurements.}
    \label{fig:schematicLMallrates}
\end{figure}

\subsection{Estimation of energy dissipation}
{The loss modulus, G$^{\prime\prime}$, responses are characterised by dissipation peaks at the initiation and termination of each temperature ramp, as seen in the insets of Figs.~\ref{fig:schematicLMallrates}(a,b). Two distinct peaks were observed in experiments performed at temperature ramp rates $>$ 1$^{\circ}$C/min. When the slowest temperature ramp of 1$^{\circ}$C/min was applied, we recorded a broad and weak peak between the points of initiation and termination of the ramp. We also observed that the loss modulus peaks became stronger and sharper at higher ramp rates. 

The peaks observed in the second derivatives of temperature, $\frac{d^{2}T}{dt^{2}}$ as reported in Figs. S3(b-d) of the supplementary material signifies intervals of rapid acceleration or deceleration and corresponds to the occurrence of sudden thermal shocks. Clearly, the larger thermal shocks at higher temperature ramp rates resulted in stronger dissipation and higher $G^{\prime\prime}$ peaks. Interestingly, researchers have used such second derivative analyses in weather forecasting, wherein the extrema in the second derivatives of the temperature denote significant and abrupt changes in thermal conditions \cite{hotz2024understanding}. For our data, the positions of the peaks in the loss modulus align with the maxima of $\frac{d^{2}T}{dt^{2}}$. The appearance of loss modulus peaks \cite{mason1996yielding, menut2012does}, therefore, can be associated with microgel rearrangement events in response to thermal shocks of varying magnitudes due to the application of temperature ramps. Rapid changes in temperature during faster heating ramps resulted in the accelerated expulsion of water from the initially jammed microgel particles. The rapid creation of free volume at faster temperature ramp rates facilitated greater energy dissipation \textit{via} enhanced rearrangements. At slower ramp rates, the particles had more time to relax, which resulted in slower energy dissipation and shorter and weaker peaks in the loss modulus response. Since rearrangements are much more difficult during the cooling ramp due to microgel swelling and subsequent jamming, cooling peaks are consistently lower, as seen in Fig.~\ref{fig:schematicLMallrates}(b). 

Figs.~\ref{fig:schematicLMallrates}(c,d) respectively display loss modulus peak heights for samples of different effective volume fractions, $\phi_{eff}$, during a fixed heating and cooling ramp rate of 20$^{\circ}$C/min. The insets show raw loss modulus data during heating and cooling of all the samples. We note stronger dissipation peaks in samples with larger effective volume fractions, except at the highest effective volume fraction, $\phi_{eff}$= 2.3. For $\phi_{eff}<$ 2, the increase in heights of the dissipation peaks with effective volume fraction implies that microgel rearrangements, driven by the release of residual stresses under the imposed thermal shock, are enhanced considerably for very high $\phi_{eff}$ values. The decrease in dissipation at $\phi_{eff}$=2.3 is a consequence of kinetic arrest. 
It has been reported earlier that the reduced number of interparticle contacts at lower effective volume fractions diminishes the restrictions imposed by neighbouring particles, leading to weaker energy dissipation \cite{ahuja2020two}. Conversely, in systems characterised by higher volume fractions, an increase in particle density enhances sample rigidity, facilitates interparticle interactions and enhances energy dissipation \cite{pham2008yielding}.  At $\phi_{eff}$= 1, suspension rigidity reduces significantly and results in noisy data as seen in Figs.~\ref{fig:schematicLMallrates}(c,d).}

Fig. S5 of the supplementary material shows an inverse correlation between $\Delta$A, the asymmetry parameter displayed earlier in Fig.~\ref{fig:schematicSMdifferentrates}, and the observed peak heights in the loss modulus response at $T_{2}$  (displayed in Fig.~\ref{fig:schematicLMallrates}).  Although this plot highlights only the peak heights of the loss modulus at temperature $T_2$, this inverse correlation applies to all the loss modulus peaks observed in our experiments. The observed inverse relationship suggests that inducing rearrangements of microgel particles can effectively eliminate path-dependent asymmetries in the sample's mechanical response during heating and cooling ramps. The Kovacs framework highlights the dependence of a glassy system's structural recovery on its thermal history \cite{bertin2003kovacs, mckenna1999tau}. The correlation observed here supports the idea that structural recovery in dense suspensions can be controlled by adjusting either the applied temperature ramp rate or the effective volume fraction of the suspension. Tuning these parameters modifies the free volume available to the microgel particles, thereby changing the frequency and strength of microgel rearrangement events. We conclude, therefore, that dissipative microgel rearrangement events can reduce asymmetries in the structural recovery of dense PNIPAM suspensions during heating and cooling processes.

\begin{figure}
    \centering
    \includegraphics[width=0.65\linewidth]{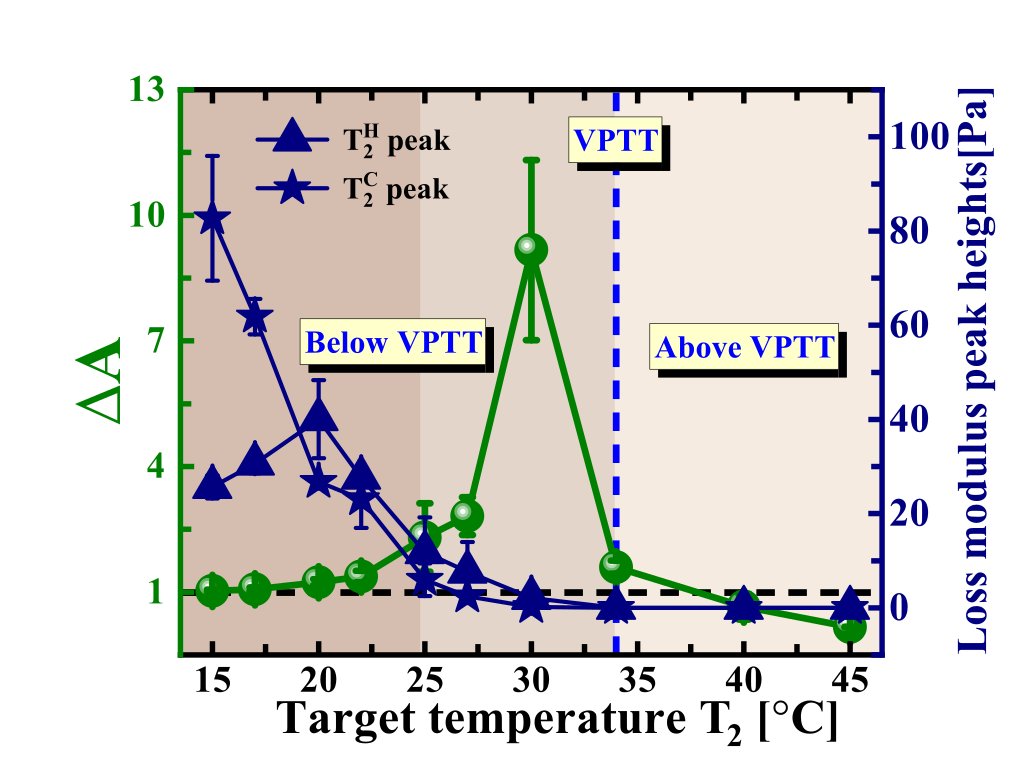}
    \caption{Asymmetry of approach, $\Delta$A, of storage modulus, G$^\prime$, of an aqueous PNIPAM suspension ($\phi_{eff}$ =1.6 at 25$^{\circ}$C) towards a target temperature $T_2$. Loss modulus peak heights, centred at $T_2$, are also plotted for both heating and cooling ramps. The experiments were performed at a constant temperature ramp rate of 10$^{\circ}$C/min for different target temperatures, $T_2$, between 10$^{\circ}$C and 45$^{\circ}$C, such that the temperature difference, $\Delta$T, was fixed at 5$^{\circ}$C. The error bars are the standard deviations calculated from three independent measurements. The temperature ranges used for the experiments are shown in Fig. S6 of the supplementary material.}
    \label{fig:schematicdifferent phieffective}
\end{figure}

\subsection{Correlating asymmetry parameter, \texorpdfstring{$\Delta$A}{DeltaA}, with loss modulus peak heights for different  \texorpdfstring{$T_2$}{T2}} 
{  The experiments were repeated at a constant ramp rate of 10$^{\circ}$C/min for different target temperatures, $T_2$, while keeping the temperature difference between the initial and target temperatures, $\Delta$T, fixed at 5$^{\circ}$C. The asymmetry of approach, $\Delta$A, of the storage modulus, G$^{\prime}$, as the system approached $T_2$ during heating and cooling experiments, and the corresponding loss modulus peak heights at $T_2$, T$^H_2$ and T$^C_2$, are plotted in Fig.~\ref{fig:schematicdifferent phieffective}. While the asymmetries increase as the volume phase transition temperature (VPTT $\approx$ 34$^{\circ}$C) is approached, we observe that the dissipative rearrangement events, as indicated by the loss modulus peak heights, decrease simultaneously. 

Figs. S6 and S7 inthe supplementary material display the raw storage and loss moduli data, respectively, for different $T_2$ values at a fixed $\Delta$T = 5$^{\circ}$C. At temperatures that are significantly lower than the VPTT, swelling-deswelling is much weaker, as indicated by the insignificant changes in $\phi_{eff}$ in Fig.~\ref{fig:schematicphieffective}(b). Under these conditions, $\Delta$T = 5$^{\circ}$C is not adequate to drive significant changes in the accessible free volume and facilitate system relaxation.

We saw from Fig.~\ref {fig:schematicphieffective}(b) that $\phi_{eff}$ decreased abruptly and monotonically due to rapid microgel deswelling within a narrow temperature range just below the VPTT. Furthermore, we reported in Fig.~\ref{fig:schematicphieffectiveAsymmetry} that significant asymmetry in the rate of change of free volume, $\Delta A_{\frac{d\phi_{\textit{eff}}{(T)}}{dt}}$, exists for experiments performed at a fixed ramp rate due to distinct pathways followed by the system during heating and cooling experiments. The rate at which free volume is created during a heating ramp is therefore very rapid when $T_2$ lies close to the VPTT. For the corresponding cooling experiment, the same system swells extremely rapidly in a process that quickly traps the available free volume. Therefore, when $T_{2}$ lies just below the VPTT, the number of pathways available to the system to approach the target temperature differs greatly in heating and cooling experiments. The corresponding asymmetries of approach towards the target temperature $T_{2}$ are therefore the highest under these conditions.

When $T_2$ lies at the VPTT of 34$^{\circ}$C (blue dashed vertical line in Fig.~\ref{fig:schematicdifferent phieffective}), the PNIPAM microgel particles exhibit maximal deswelling, and very few microgel suspension structures survive under these conditions \cite{misra2024effect}. Consequently, the approach of G$^{\prime}$ towards T$_2$ is symmetric due to the removal of all dynamical constraints. As the temperature $T_2$ is increased beyond the VPTT, the collapsed particles become increasingly hydrophobic and self-assemble to form gel networks \cite{romeo2010temperature}. Under these conditions and as shown in Fig.~\ref {fig:schematicphieffective}(b), changes in $\phi_{eff}$ due to temperature variations during heating and cooling are insignificant when compared to the scenario below the VPTT. Formation of a collapsed gel state, held together by strong temperature-dependent hydrophobic attractions \cite{romeo2010temperature}, precludes particle rearrangement events and removes path-dependent transitions. Dissipation and asymmetry are therefore minimal above the VPTT, and an inverse correlation between asymmetry and loss modulus peak height as reported below the VPTT does not exist. The results reported here, therefore, highlight important differences in the structural recovery of glasses and gels.} 

 In all the temperature ramp experiments discussed above, the asymmetry of approach of G$^{\prime}$ towards the target temperature $T_{2}$, together with the appearance of loss modulus peaks at the initiation and termination of the temperature ramps, is interpreted as signatures of the complex structural recovery of dense PNIPAM suspensions in response to external driving forces (temperature ramps in the present experiments). A comparative analysis of these relaxation characteristics with those observed in Newtonian glycerol, assessed during a temperature ramp experiment conducted at 20$^{\circ}$C/min, reveals that G$^{\prime}$ is negligible (Fig.S8(b) of the supplementary material) and loss modulus peaks are absent (Fig.S8(c)) at the initiation and termination of the temperature ramps. The path-dependent relaxation of dense colloidal PNIPAM systems is therefore a consequence of their disordered and metastable nature. Our results clearly underscore the distinct differences in the relaxations of glasses and Newtonian materials.

\section{Conclusions}
{According to the energy landscape paradigm \cite{stillinger2002energy, richert2002heterogeneous}, a disordered and metastable material driven towards thermal equilibrium explores many potential energy wells in its free energy landscape as it navigates towards its free energy minimum.  In contrast, dense packings within the glassy regime lower the mobility of the system, trapping the system in a potential energy well that does not represent the lowest free energy state. This theoretical framework results in the sensitive dependence of structural recovery on the thermal history of the experiment, rather than only on the final and initial thermal states \cite{mckenna201750th}. The application
of temperature ramps of varying magnitudes to dense colloidal suspensions of thermoresponsive,
deformable and compressible microgel particles should serve as an effective strategy to probe the
nonlinear path-dependent structural recovery of these systems and to tailor their memories.

We investigated the complex structural recovery of dense colloidal suspensions by applying temperature ramps at different rates. The use of thermoresponsive PNIPAM particles in dense suspensions allowed us to mimic the dynamical slowing down of a supercooled liquid under a temperature quench \cite{romeo2010temperature, misra2024effect}. As a result of the thermal shock to the system at the initiation and termination of the temperature ramp, the suspension storage modulus approached the target temperature asymmetrically at low temperature ramp rates, while such asymmetry disappeared as the ramp rate was increased. The loss modulus response displayed distinct peaks at the initiation and termination of the temperature ramp, with the strengths of the peaks seen to increase with increasing temperature ramp rates. We associated the asymmetries in the storage modulus data under heating and cooling temperature ramps with the path-dependent structural recovery of the suspension as it navigated a complex, thermal history-dependent potential energy landscape \cite{mckenna201750th}. Interestingly, we uncovered an inverse correlation between the asymmetry of approach of the storage modulus towards the target temperature and the peak heights in the loss modulus responses. Our data clearly reveals that the application of a strong thermal shock can effectively smoothen the potential energy landscape \textit{via} particle rearrangement events. Higher ramp rates are more effective in driving cooperative microgel rearrangement events and removing asymmetries in the elastic response of the system in heating and cooling experiments. 

We also performed temperature ramp experiments at a fixed ramp rate with dense PNIPAM suspensions of different effective volume fractions. These experiments verified our previous results and revealed that path-dependent asymmetries in jammed systems depend sensitively on the accessible free volume and can be removed by inducing rearrangement events \textit{via} external drives (thermal shocks in our experiments). The asymmetries observed in the approach of the storage modulus towards a target temperature below the VPTT and the presence of dissipation peaks \cite{mason1996yielding} due to the onset of rearrangement cascades are signatures of the complex relaxation dynamics of driven glassy systems. In the future, it will be intriguing to investigate how the stiffness of microgel particles, synthesised with varying concentrations of crosslinkers, affects the relaxation dynamics of their dense suspensions under heating and cooling temperature ramps. Understanding the interplay between the structural recovery of jammed systems and the particle-scale dissipation dynamics will enhance our knowledge of non-equilibrium relaxation processes.

Application of temperature ramps to dense, metastable suspensions and the investigation of their viscoelastic responses therefore provide profound insights into the study of the intricate path-dependent dynamics and the role of dissipative processes in the relaxation behaviours of driven complex materials. Our study offers a novel route to systematically tailor the memory and dissipation in soft glassy materials. It suggests a framework for preparing adaptive systems with programmable thermal histories, and can be extended to the study of nonequilibrium relaxation pathways in a range of stimuli-responsive systems. Our research is important for the development of soft robots and novel biomaterials that are resilient, intelligent, and multifunctional.}

\newpage
\section*{ACKNOWLEDGEMENTS}
The authors acknowledge Sayantan Majumdar and Rituparno Mandal for useful discussions and valuable suggestions, and Vaibhav Raj Singh Parmar for his useful comments on the manuscript.

\section*{Conflicts of interest} 
The authors have no conflict of interest to disclose.

\section*{Data availability}
Data will be made available on request.

\renewcommand\refname{References}
	\bibliographystyle{elsarticle-num}
	\bibliography{Reference}

\begin{thebibliography}{10}
\expandafter\ifx\csname url\endcsname\relax
  \def\url#1{\texttt{#1}}\fi
\expandafter\ifx\csname urlprefix\endcsname\relax\def\urlprefix{URL }\fi
\expandafter\ifx\csname href\endcsname\relax
  \def\href#1#2{#2} \def\path#1{#1}\fi

\bibitem{pusey1986phase}
P.~N. Pusey, W.~Van~Megen, Phase behaviour of concentrated suspensions of nearly hard colloidal spheres, Nature 320~(6060) (1986) 340--342.

\bibitem{mattsson2009soft}
J.~Mattsson, H.~M. Wyss, A.~Fernandez-Nieves, K.~Miyazaki, Z.~Hu, D.~R. Reichman, D.~A. Weitz, Soft colloids make strong glasses, Nature 462~(7269) (2009) 83--86.

\bibitem{di2011signatures}
X.~Di, K.~Win, G.~B. McKenna, T.~Narita, F.~Lequeux, S.~R. Pullela, Z.~Cheng, Signatures of structural recovery in colloidal glasses, Physical review letters 106~(9) (2011) 095701.

\bibitem{vlassopoulos2014tunable}
D.~Vlassopoulos, M.~Cloitre, Tunable rheology of dense soft deformable colloids, Current opinion in colloid \& interface science 19~(6) (2014) 561--574.

\bibitem{basak2013encapsulation}
R.~Basak, R.~Bandyopadhyay, Encapsulation of hydrophobic drugs in pluronic f127 micelles: effects of drug hydrophobicity, solution temperature, and ph, Langmuir 29~(13) (2013) 4350--4356.

\bibitem{negi2014viscoelasticity}
A.~S. Negi, C.~G. Redmon, S.~Ramakrishnan, C.~O. Osuji, Viscoelasticity of a colloidal gel during dynamical arrest: Evolution through the critical gel and comparison with a soft colloidal glass, Journal of Rheology 58~(5) (2014) 1557--1579.

\bibitem{suman2020universality}
K.~Suman, Y.~M. Joshi, On the universality of the scaling relations during sol-gel transition, Journal of Rheology 64~(4) (2020) 863--877.

\bibitem{suman2021rheological}
K.~Suman, S.~Sourav, Y.~M. Joshi, Rheological signatures of gel--glass transition and a revised phase diagram of an aqueous triblock copolymer solution of pluronic f127, Physics of Fluids 33~(7) (2021).

\bibitem{suman2022anomalous}
K.~Suman, N.~J. Wagner, Anomalous rheological aging of a model thermoreversible colloidal gel following a thermal quench, The Journal of Chemical Physics 157~(2) (2022).

\bibitem{dufresne2004preparation}
M.-H. Dufresne, D.~Le~Garrec, V.~Sant, J.-C. Leroux, M.~Ranger, Preparation and characterization of water-soluble ph-sensitive nanocarriers for drug delivery, International journal of pharmaceutics 277~(1-2) (2004) 81--90.

\bibitem{romeo2010temperature}
G.~Romeo, A.~Fernandez-Nieves, H.~M. Wyss, D.~Acierno, D.~A. Weitz, Temperature-controlled transitions between glass, liquid, and gel states in dense p-nipa suspensions, Advanced Materials 22~(31) (2010) 3441.

\bibitem{franco2021glass}
S.~Franco, E.~Buratti, V.~Nigro, E.~Zaccarelli, B.~Ruzicka, R.~Angelini, Glass and jamming rheology in soft particles made of pnipam and polyacrylic acid, International journal of molecular sciences 22~(8) (2021) 4032.

\bibitem{wu2003phase}
J.~Wu, B.~Zhou, Z.~Hu, Phase behavior of thermally responsive microgel colloids, Physical review letters 90~(4) (2003) 048304.

\bibitem{lyon2012polymer}
L.~A. Lyon, A.~Fernandez-Nieves, The polymer/colloid duality of microgel suspensions, Annual review of physical chemistry 63~(1) (2012) 25--43.

\bibitem{karg2008temperature}
M.~Karg, I.~Pastoriza-Santos, B.~Rodriguez-Gonzalez, R.~Von~Klitzing, S.~Wellert, T.~Hellweg, Temperature, ph, and ionic strength induced changes of the swelling behavior of pnipam- poly (allylacetic acid) copolymer microgels, Langmuir 24~(12) (2008) 6300--6306.

\bibitem{peppas2006hydrogels}
N.~A. Peppas, J.~Z. Hilt, A.~Khademhosseini, R.~Langer, Hydrogels in biology and medicine: from molecular principles to bionanotechnology, Advanced materials 18~(11) (2006) 1345--1360.

\bibitem{misra2020influence}
C.~Misra, S.~K. Behera, R.~Bandyopadhyay, Influence of particle size on the thermoresponsive and rheological properties of aqueous poly (n-isopropylacrylamide) colloidal suspensions, Bulletin of Materials Science 43~(1) (2020) 182.

\bibitem{vialetto2024effect}
J.~Vialetto, S.~N. Ramakrishna, L.~Isa, M.~Laurati, Effect of particle stiffness and surface properties on the non-linear viscoelasticity of dense microgel suspensions, Journal of Colloid and Interface Science 672 (2024) 814--823.

\bibitem{gury2025internal}
L.~Gury, M.~Gauthier, J.-M. Suau, D.~Vlassopoulos, M.~Cloitre, Internal microstructure dictates yielding and flow of jammed suspensions and emulsions, ACS nano 19~(15) (2025) 14931--14940.

\bibitem{zhang2025unveiling}
L.~Zhang, J.~Cao, P.~W.~F. Yeung, H.~Jiang, J.~Jiang, W.~Liu, T.~Ngai, Unveiling the rheological behavior of poly (styrene-co-n-isopropylacrylamide) microgel aqueous suspensions, Journal of Colloid and Interface Science 685 (2025) 1173--1183.

\bibitem{franco2025soft}
S.~Franco, B.~Ruzicka, R.~Angelini, Soft and responsive: rheological insights into pnipam based microgels and applications, Journal of Physics: Condensed Matter 37~(24) (2025) 243001.

\bibitem{misra2024effect}
C.~Misra, S.~V. Kawale, S.~K. Behera, R.~Bandyopadhyay, Effect of particle stiffness on microgel self-assembly and suspension phase behavior over a broad temperature range, Physics of Fluids 36~(10) (2024).

\bibitem{10.1007/BFb0050366}
A.~J. Kovacs, Transition vitreuse dans les polym{\`e}res amorphes. etude ph{\'e}nom{\'e}nologique, in: Fortschritte Der Hochpolymeren-Forschung, Springer Berlin Heidelberg, Berlin, Heidelberg, 1964, pp. 394--507.

\bibitem{tool1946viscosity}
A.~Q. Tool, Viscosity and the extraordinary heat effects in glass, J. Am. Ceram. Soc 29~(9) (1946) 240--253.

\bibitem{struik1978physical}
L.~C.~E. Struik, et~al., Physical aging in amorphous polymers and other materials, Vol. 106, Citeseer, 1978.

\bibitem{larson1999structure}
R.~G. Larson, The structure and rheology of complex fluids, (No Title) (1999).

\bibitem{bertin2003kovacs}
E.~Bertin, J.-P. Bouchaud, J.~Drouffe, C.~Godreche, The kovacs effect in model glasses, Journal of physics A: mathematical and general 36~(43) (2003) 10701.

\bibitem{yunker2009irreversible}
P.~Yunker, Z.~Zhang, K.~B. Aptowicz, A.~G. Yodh, Irreversible rearrangements, correlated domains, and local structure in aging glasses, Physical review letters 103~(11) (2009) 115701.

\bibitem{joshi2014long}
Y.~M. Joshi, Long time response of aging glassy polymers, Rheologica Acta 53~(5) (2014) 477--488.

\bibitem{scalliet2019rejuvenation}
C.~Scalliet, L.~Berthier, Rejuvenation and memory effects in a structural glass, Physical review letters 122~(25) (2019) 255502.

\bibitem{edera2025mechanical}
P.~Edera, M.~Bantawa, S.~Aime, R.~T. Bonnecaze, M.~Cloitre, Mechanical tuning of residual stress, memory, and aging in soft glassy materials, Physical Review X 15~(1) (2025) 011043.

\bibitem{mckenna1999tau}
G.~McKenna, M.~Vangel, A.~L. Rukhin, S.~D. Leigh, B.~Lotz, C.~Straupe, The $\tau$-effective paradox revisited: an extended analysis of kovacs’ volume recovery data on poly (vinyl acetate), Polymer 40~(18) (1999) 5183--5205.

\bibitem{lira2021fundamental}
J.~Lira-Escobedo, P.~Mendoza-M{\'e}ndez, M.~Medina-Noyola, G.~McKenna, P.~Ram{\'\i}rez-Gonz{\'a}lez, On a fundamental description of the kovacs’ kinetic signatures in glass-forming systems, The Journal of Chemical Physics 155~(1) (2021).

\bibitem{mckenna201750th}
G.~B. McKenna, S.~L. Simon, 50th anniversary perspective: Challenges in the dynamics and kinetics of glass-forming polymers, Macromolecules 50~(17) (2017) 6333--6361.

\bibitem{kaushal2016analyzing}
M.~Kaushal, Y.~M. Joshi, Analyzing aging under oscillatory strain field through the soft glassy rheology model, The Journal of Chemical Physics 144~(24) (2016).

\bibitem{shukla2017boltzmann}
A.~Shukla, Y.~M. Joshi, Boltzmann superposition principle for a time-dependent soft material: assessment under creep flow field, Rheologica Acta 56~(11) (2017) 927--940.

\bibitem{agarwal2020signatures}
M.~Agarwal, M.~Kaushal, Y.~M. Joshi, Signatures of overaging in an aqueous dispersion of carbopol, Langmuir 36~(48) (2020) 14849--14863.

\bibitem{di2014dynamics}
X.~Di, X.~Peng, G.~B. McKenna, Dynamics of a thermo-responsive microgel colloid near to the glass transition, The Journal of chemical physics 140~(5) (2014).

\bibitem{peng2016physical}
X.~Peng, G.~B. McKenna, Physical aging and structural recovery in a colloidal glass subjected to volume-fraction jump conditions, Physical Review E 93~(4) (2016) 042603.

\bibitem{Dhavale}
T.~P. Dhavale, S.~Jatav, Y.~M. Joshi, Thermally activated asymmetric structural recovery in a soft glassy nano-clay suspension, Soft Matter 9 (2013) 7751--7756.

\bibitem{robin2022glass}
C.~Robin, C.~G. Robertson, Glass-like signatures in the dynamic rheology of particle-filled polymers, Macromolecules 55~(7) (2022) 2729--2738.

\bibitem{chen2023memory}
Y.~Chen, Q.~Zhang, S.~Ramakrishnan, R.~L. Leheny, Memory in aging colloidal gels with time-varying attraction, The Journal of chemical physics 158~(2) (2023).

\bibitem{chen2010theory}
K.~Chen, K.~S. Schweizer, Theory of aging, rejuvenation, and the nonequilibrium steady state in deformed polymer glasses, Physical Review E—Statistical, Nonlinear, and Soft Matter Physics 82~(4) (2010) 041804.

\bibitem{janzen2024rejuvenation}
G.~Janzen, L.~M. Janssen, Rejuvenation and memory effects in active glasses induced by thermal and active cycling, Physical Review Research 6~(2) (2024) 023257.

\bibitem{mcphee1993poly}
W.~McPhee, K.~C. Tam, R.~Pelton, Poly (n-isopropylacrylamide) latices prepared with sodium dodecyl sulfate, Journal of colloid and interface science 156~(1) (1993) 24--30.

\bibitem{carrier2009nonlinear}
V.~Carrier, G.~Petekidis, Nonlinear rheology of colloidal glasses of soft thermosensitive microgel particles, Journal of Rheology 53~(2) (2009) 245--273.

\bibitem{macosko1994rheology}
C.~W. Macosko, Rheology principles, Measurements and Applications (1994).

\bibitem{wang2015asymmetric}
M.~Wang, K.~Zhang, Z.~Li, Y.~Liu, J.~Schroers, M.~D. Shattuck, C.~S. O'Hern, Asymmetric crystallization during cooling and heating in model glass-forming systems, Physical Review E 91~(3) (2015) 032309.

\bibitem{viollaz2003structural}
P.~Viollaz, S.~Alzamora, A.~Nieto, Structural relaxation in glasses: numerical exploration of variables of narayanaswamy and moynihan's model, Journal of food engineering 56~(4) (2003) 393--399.

\bibitem{ibanez2024heating}
M.~Ib{\'a}nez, C.~Dieball, A.~Lasanta, A.~Godec, R.~A. Rica, Heating and cooling are fundamentally asymmetric and evolve along distinct pathways, Nature Physics 20~(1) (2024) 135--141.

\bibitem{stillinger2002energy}
F.~H. Stillinger, P.~G. Debenedetti, Energy landscape diversity and supercooled liquid properties, The Journal of chemical physics 116~(8) (2002) 3353--3361.

\bibitem{hornig2012static}
R.~Hornig, J.~Sunder, B.~Herr, Static cold sealing force behaviour of amorphous hnbr materials, International Polymer Science and Technology 39~(11) (2012) 1--14.

\bibitem{hotz2024understanding}
B.~Hotz, L.~Papritz, M.~R{\"o}thlisberger, Understanding the vertical temperature structure of recent record-shattering heatwaves, Weather and Climate Dynamics 5~(1) (2024) 323--343.

\bibitem{mason1996yielding}
T.~Mason, J.~Bibette, D.~Weitz, Yielding and flow of monodisperse emulsions, Journal of colloid and interface science 179~(2) (1996) 439--448.

\bibitem{menut2012does}
P.~Menut, S.~Seiffert, J.~Sprakel, D.~A. Weitz, Does size matter? elasticity of compressed suspensions of colloidal-and granular-scale microgels, Soft Matter 8~(1) (2012) 156--164.

\bibitem{ahuja2020two}
A.~Ahuja, A.~Potanin, Y.~M. Joshi, Two step yielding in soft materials, Advances in Colloid and Interface Science 282 (2020) 102179.

\bibitem{pham2008yielding}
K.~Pham, G.~Petekidis, D.~Vlassopoulos, S.~Egelhaaf, W.~Poon, P.~Pusey, Yielding behavior of repulsion-and attraction-dominated colloidal glasses, Journal of Rheology 52~(2) (2008) 649--676.

\bibitem{richert2002heterogeneous}
R.~Richert, Heterogeneous dynamics in liquids: fluctuations inspace and time, Journal of Physics: Condensed Matter 14~(23) (2002) R703.

\end{thebibliography}
    

\section*{Supplementary material}

# 1 Measurements of average hydrodynamic diameters and temperature-dependent swelling ratios of PNIPAM microgel particles in aqueous suspension

Dynamic light scattering (DLS) experiments were conducted using a Brookhaven Instruments Corporation (BIC) BI-200SM spectrometer. More details about the setup can be found elsewhere [1] [2]. A Polyscience digital water circulating temperature controller was used to maintain the sample temperature within the range 12°C and 55°C. The intensity autocorrelation function, $g^{(2)}(q,t)$, of the scattered light was calculated using a Brookhaven Instruments Corporation digital autocorrelator, BI-9000AT. Here, $g^{(2)}(q,t) = \frac{\langle I(q,0)I(q,t)\rangle}{\langle I(q,0)\rangle^2} = 1 + A|g^{(1)}(q,t)|^2$ [3], where $I(q,t)$, $g^{(1)}(q,t)$ and A are the scattered intensity at a particular scattering wave vector $q$ and a delay time t, the normalized electric field autocorrelation function and the coherence factor, respectively. The angular bracket $\langle \rangle$ in the expression represents an average over time. The scattering wave vector $q = (4\pi n/\lambda)\sin(\theta/2)$, where $\theta$, n and $\lambda$ are, respectively, the scattering angle (60° used here), the refractive index of the medium ($\approx$1.33) and the laser wavelength (532 nm). For a suspension of particles characterised by size polydispersity, the decay of the normalised autocorrelation function, $C(t) = \frac{g^{(2)}(q,t)-1}{A}$ [4], is fitted to a stretched exponential function of the form $C(t) = [e^{-(t/\tau)^\beta}]^2$. The mean relaxation time of the suspended particles (the scatterers) is estimated using the relation $\langle\tau\rangle = (\frac{\tau}{\beta})\Gamma(\frac{1}{\beta})$, where $\beta$ is the stretching exponent and $\Gamma$ is the Euler Gamma function [5]. The average hydrodynamic diameter of the particles, $\langle d_H \rangle$, is estimated using the Stokes-Einstein relation $\langle d_H \rangle = \frac{k_B T \langle\tau\rangle q^2}{3\pi\eta}$, where $k_B$, $T$, and $\eta$ are the Boltzmann constant, the absolute temperature, and the viscosity of the solvent, respectively [4] [5]. The error bars were calculated from three measurements. Fig. S1 shows the temperature-dependent swelling ratios of PNIPAM particles. The inset shows representative normalised intensity autocorrelation data at 25°C and the corresponding fit to a stretched exponential function. For the measurement of hydrodynamic diameter, we waited 20 minutes at each temperature to ensure equilibration.

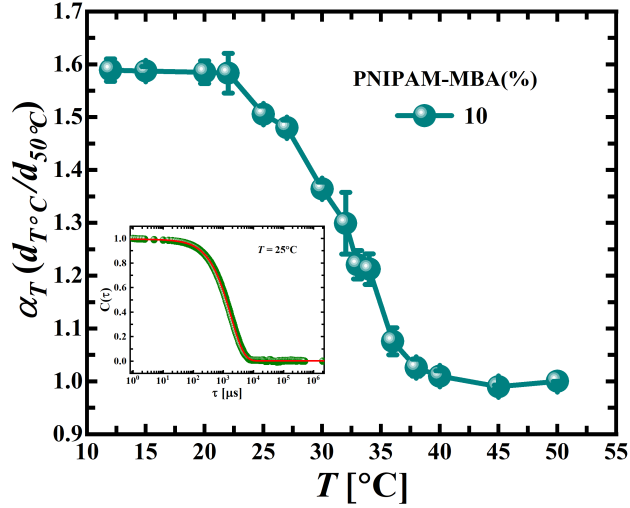

Fig. S1: (a) Temperature-dependent swelling ratios, $\alpha_T = (d_{T°C}/d_{50°C})$, of PNIPAM particles in aqueous suspension. Here, $d_{T°C}$ and $d_{50°C}$ are the average hydrodynamic diameters of PNIPAM particles at temperatures $T$ and 50°C, respectively. The inset shows a representative plot of a normalised intensity autocorrelation function (green circles) at 25°C, plotted versus delay time. The red line is a stretched exponential fit.

## 2  Estimation of effective volume fraction

The viscosity, $\eta$, of aqueous suspensions of PNIPAM was measured by rotational rheometry while varying the shear rates, $\dot{\gamma}$, from 0.01 to 1000 $s^{-1}$. The zero-shear viscosities, $\eta_0$, were estimated by fitting the viscosity vs. shear rate, $\dot{\gamma}$, data to the Cross model [6]:

$$\frac{(\eta - \eta_\infty)}{(\eta_0 - \eta_\infty)} = \frac{1}{1 + (k\dot{\gamma})^m} \tag{1}$$

In Eqn. 1, $\eta_0$ and $\eta_\infty$ are the low and high shear rate viscosities, respectively, $k$ is a time constant that corresponds to the relaxation time of the microgel particles in suspension, and $m$ is a dimensionless exponent. The inset in Fig. S2 shows $\eta$ versus $\dot{\gamma}$ plots and the corresponding fits to the Cross model with fitting parameters $\eta_0$, $\eta_\infty$, $k$ and $m$. The relative viscosity, $\eta_{rel}$, of an aqueous suspension of PNIPAM is given by the ratio of the zero-shear viscosity, $\eta_0$, of each suspension, and the viscosity of water, $\eta_s$ (0.89 mPa.s at 25°C). Thus, $\eta_{rel} = \eta_0/\eta_s$. Since the volume fraction, $\phi$, is inappropriate for quantifying the packing of deformable and compressible (soft) particles in suspension, we used a modified parameter known as the effective volume fraction, $\phi_{eff} = nV_d$, [7–9], where $n$ represents particle number density per unit volume, and $V_d = \frac{\pi(\langle d_H \rangle)^3}{6}$ is the volume of an undeformed particle

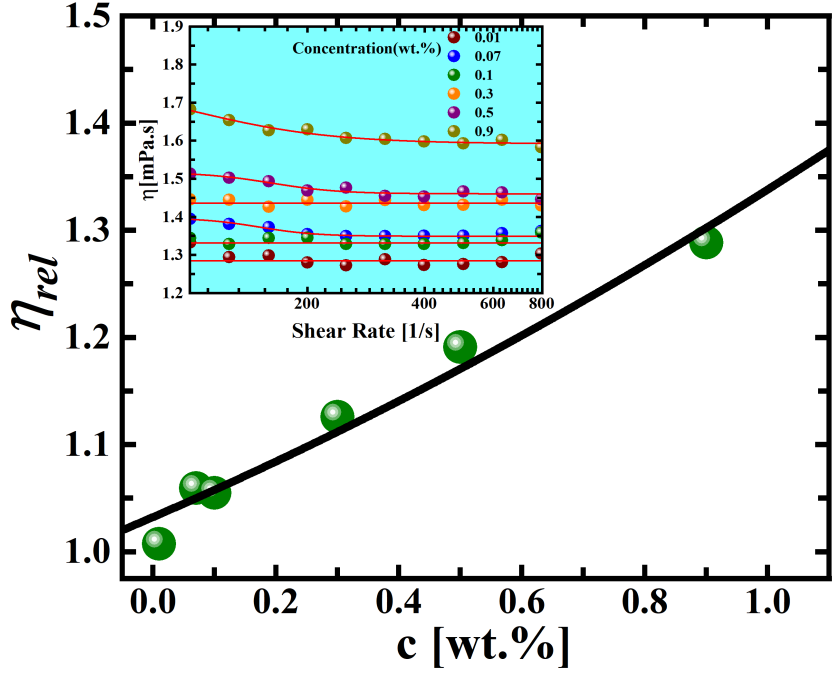

Fig. S2: Relative viscosity, $\eta_{rel}$, vs. particle concentration, c in wt.%, of aqueous suspensions of PNIPAM particles. The black solid line is a fit to Eqn. 2 of Supplementary material. The inset shows $\eta$ versus $\dot{\gamma}$ plots for the same suspensions and the corresponding fits to the Cross model (Eqn. 1 of Supplementary material).

with an average hydrodynamic diameter $\langle d_H \rangle$ in a dilute suspension. The relative viscosities, $\eta_{rel}$, estimated at 25°C, are plotted versus particle concentration, c, in Fig. S2 and are fitted to the Batchelor equation [10, 11]:

$$\eta_{rel} = 1 + 2.5(c/c_p) + 5.9(c/c_p)^2 \qquad (2)$$

The polymer concentration inside each particle, $c_p$, is extrapolated from these fits (black line in Fig. S2). The effective volume fractions are next calculated using the relation $\phi_{eff} = c/c_p$. The temperature-dependent effective volume fractions, calculated from the hydrodynamic diameter data using Eqn. 1 of the main manuscript, are plotted in Fig. S4(a).

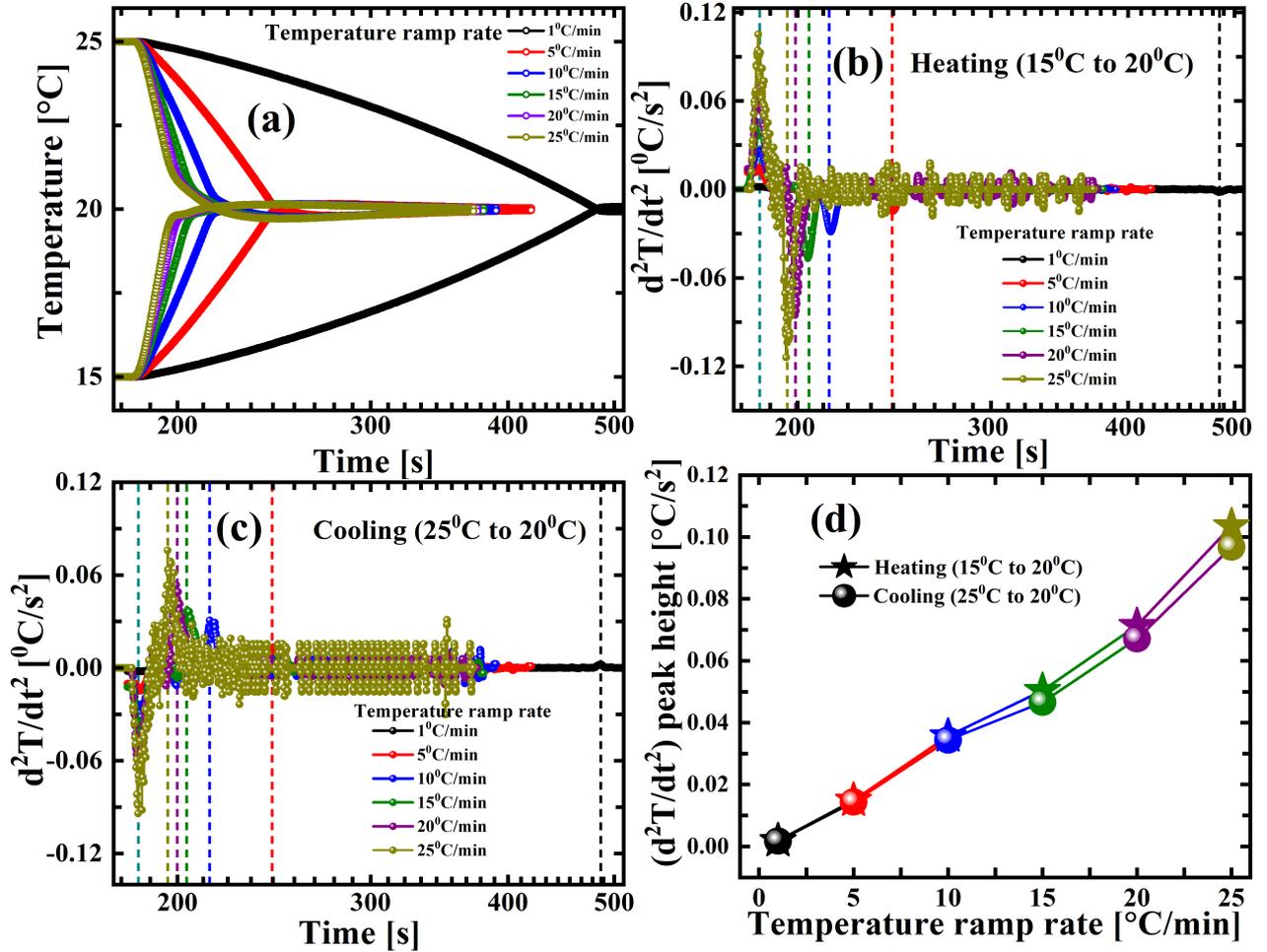

Fig. S3: (a) Temperature *vs.* time profiles in heating and cooling temperature ramp experiments. The second derivatives of temperature with time, $\frac{d^2T}{dt^2}$, during (b) heating and (c) cooling ramps. The short dashed vertical lines indicate the positions at which the second derivatives are maximum for different ramp rates. (d) Maximum magnitudes of the second derivatives at the initiation of heating and cooling temperature ramps. The increasing trend is reproduced during both ramp initiation and termination.

# 3 Estimation of asymmetric approach of effective volume fraction towards target temperature for different temperature ramp rates

We estimated temporal changes in the effective volume fractions of PNIPAM suspensions (of $\phi_{eff}$ = 1.6 at 25°C) that were heated and cooled to a target temperature $T_2 = 20$°C at different temperature ramp rates. The initial temperatures for the heating and cooling ramp experiments were 15°C and 25°C respectively. The temperature change $\Delta T$ was therefore kept fixed at 5°C.

Temperature-dependent effective volume fraction data, $\phi_{eff}(T)$, calculated using Eqn. 1 of the main manuscript, was fitted to an exponential function of the form $e^{-(T/T^*)^\beta}$ ($T$ in °C) [12], and is represented by the red line in Fig. S4(a). Here $T^*$ and $\beta$ are fitting parameters. We ignored the deformability and compressibility of the microgels and considered them to be hard spheres. The value of $\phi_{eff}$ at each temperature was determined through interpolation of the fitted data presented in Fig. S4(a). We next extracted $\phi_{eff}(T)$ values separately at every instant for all the heating and cooling temperature ramp experiments reported here.

Fig. S4(b) illustrates the temporal variations of $\phi_{eff}(T)$, interpolated according to the procedure discussed above, during heating and cooling experiments at ramp rates between 1°C/min and 25°C/min. The asymmetry parameter, $\Delta A_{\phi_{eff}}$, plotted in Fig. S4(c), was estimated as the ratio of the areas enclosed by the $\phi_{eff}(T)$-time curves during heating and cooling processes at different applied temperature ramp rates. The insets of Fig. S4(c) highlight these areas of interest (green for the heating ramp and blue for the cooling ramp). With increase in applied temperature ramp rate, we noted a decrease in the asymmetry of approach of $\phi_{eff}(T)$ towards the target temperature $T_2$. Next, we calculated the rate of change of $\phi_{eff}(T)$ with time, $\frac{d\phi_{eff}(T)}{dt}$, by taking time-derivatives using a 3-point centred difference method in Origin software for the data presented in Fig. S4(b). The results of these calculations for heating and cooling experiments at all the temperature ramp rates are presented in Fig. S4(d).

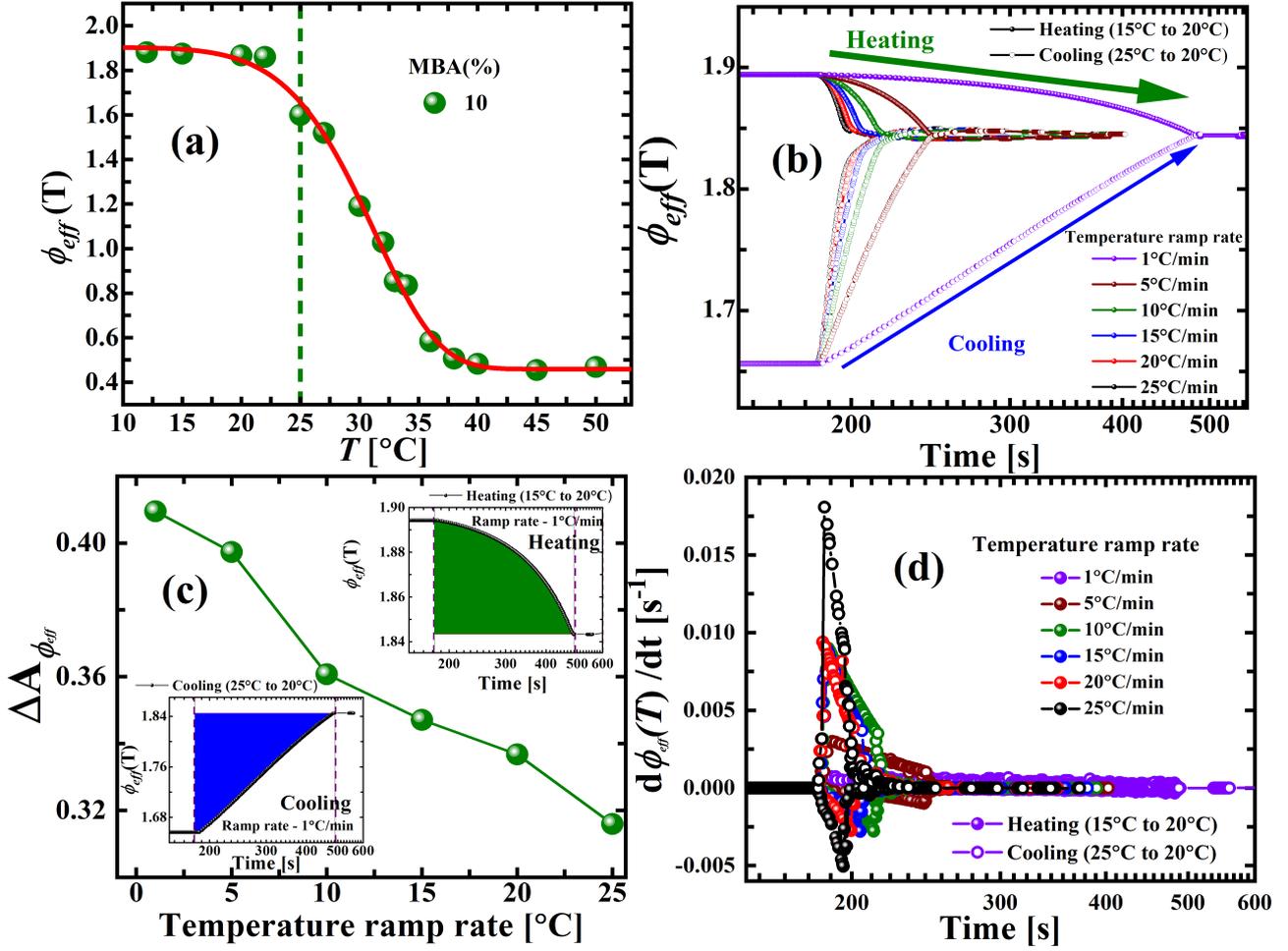

Fig. S4: (a) Temperature-dependent effective volume fraction, $\phi_{eff}(T)$, of a PNIPAM suspension prepared at $\phi_{eff}$ =1.6 at 25°C and the corresponding fit to an exponential function (red line). (b) Effective volume fraction vs. time during heating and cooling processes at ramp rates between 1°C/min and 25°C/min. (c) The asymmetric approach of $\phi_{eff}$, $\Delta A_{\phi_{eff}}$, towards the target temperature under heating and cooling temperature ramps, applied at rates between 1°C/min and 25°C/min. Insets show the temporal changes in effective volume fraction during heating (15°C to 20°C) and cooling (25°C to 20°C) ramps. The green and blue shaded regions indicate the areas enclosed by the $\phi_{eff}$-time curves during the heating and cooling processes, respectively. (b) Rate of change of effective volume fraction, $\frac{d\phi_{eff}(T)}{dt}$, vs. time during heating and cooling processes at ramp rates between 1°C/min and 25°C/min.

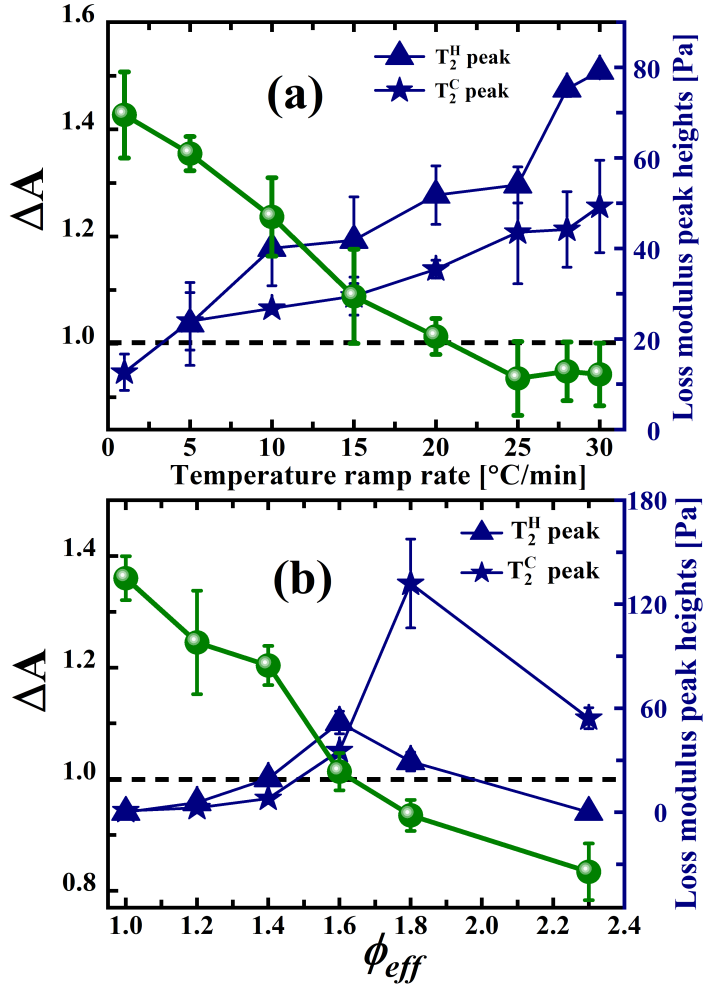

Fig. S5: Inverse correlation between the asymmetry of approach, $\Delta A$, of storage modulus, $G'$, towards the target temperature $T_2 = 20°C$, and the loss modulus peak heights at $T_2$ for (a) different heating and cooling ramp rates applied to microgel suspensions prepared at $\phi_{eff} = 1.6$ at 25°C and (b) different effective volume fractions of PNIPAM microgel suspensions under a constant applied temperature ramp of 20°C/min.

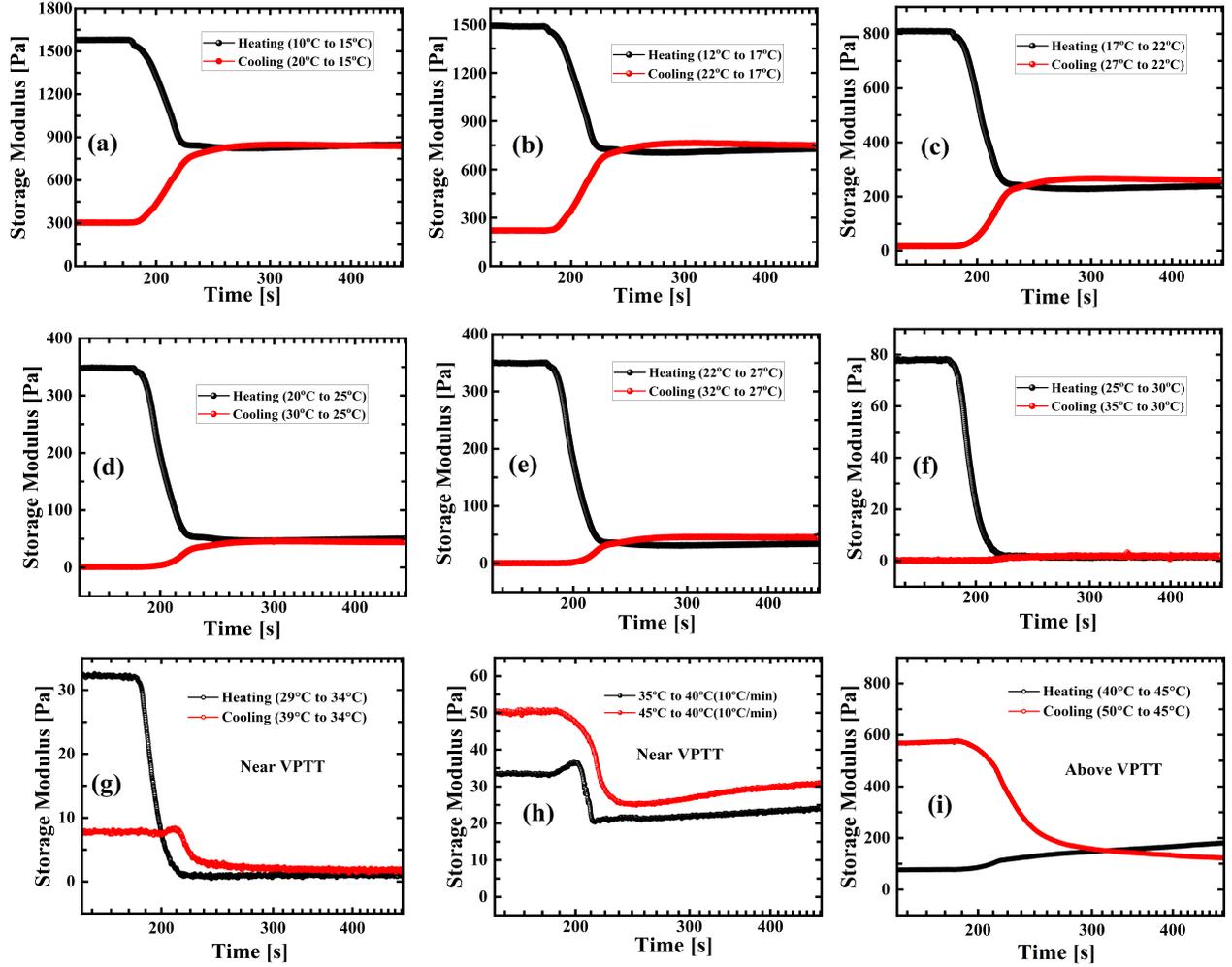

Fig. S6: Storage modulus responses during heating and cooling ramps for temperature changes ($T_1$ to $T_2$ and $T_3$ to $T_2$) between (a) 10°C to 15°C and 20°C to 15°C, (b) 12°C to 17°C and 22°C to 17°C, (c) 17°C to 22°C and 27°C to 22°C, (d) 20°C to 25°C and 30°C to 25°C, (e) 22°C to 27°C and 32°C to 27°C, (f) 25°C to 30°C and 35°C to 30°C, (g) 29°C to 34°C and 39°C to 34°C, (h) 35°C to 40°C and 45°C to 40°C, (i) 40°C to 45°C and 50°C to 45°C, applied to an aqueous PNIPAM suspension prepared at effective volume fraction, $\phi_{eff} = 1.6$, at 25°C. All measurements were performed at a constant temperature ramp rate of 10°C/min and $\Delta T = 5$°C.

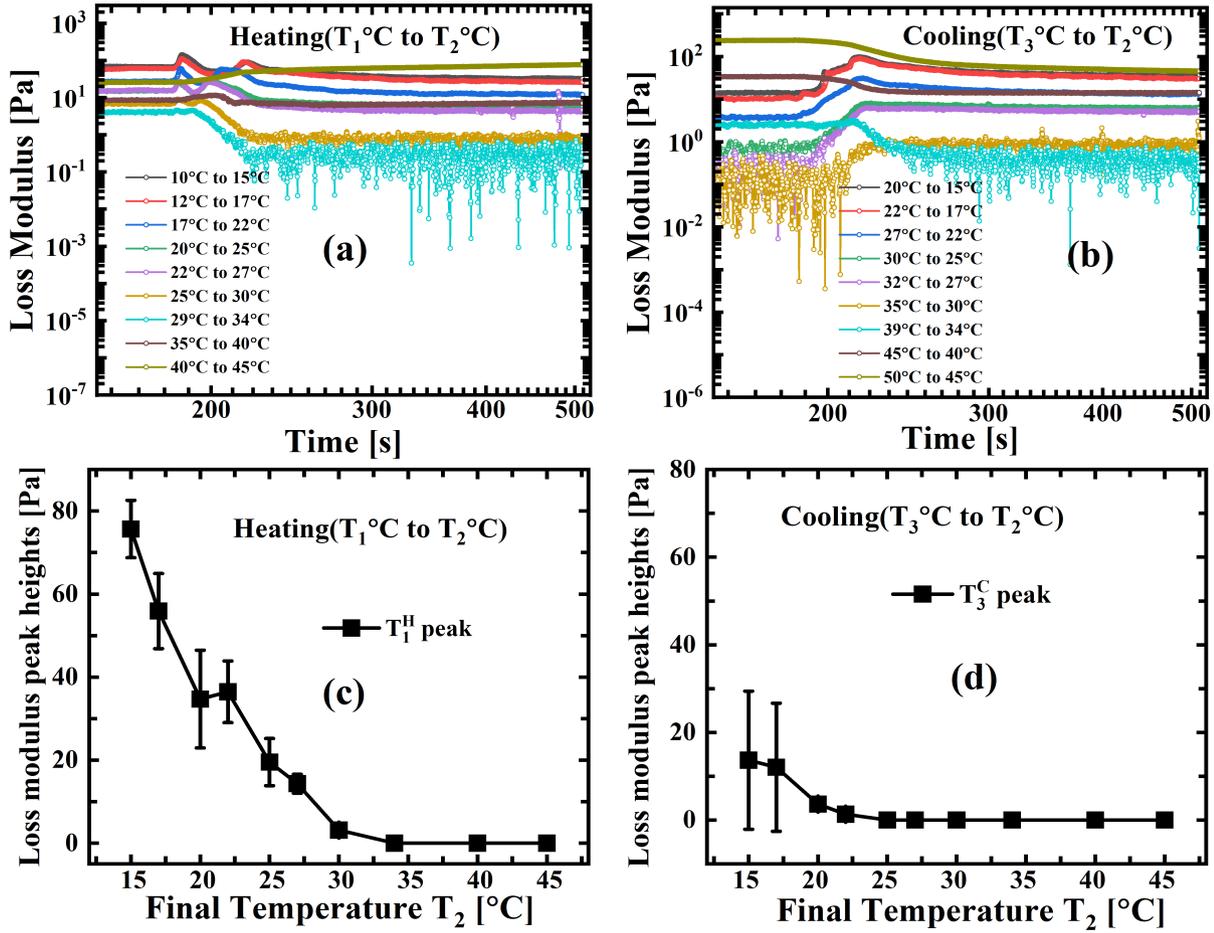

Fig. S7: Loss modulus relaxation in the same experiments as in Fig. S6 during (a) heating and (b) cooling at a fixed $\Delta T = 5°C$. Loss modulus peak heights during (c) heating and (d) cooling processes. All measurements were performed at a constant temperature ramp rate of 10°C/min.

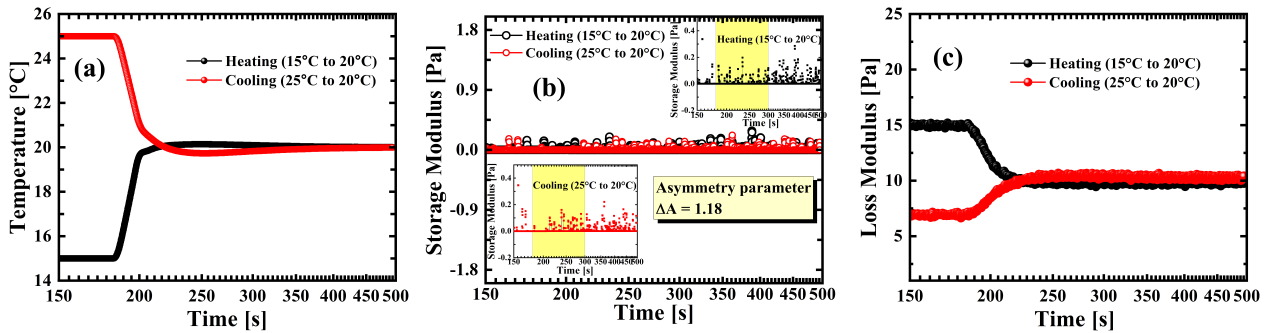

Fig. S8: (a) Applied temperature profile, and the temporal evolutions of (b) storage modulus, $G'$, and (c) loss modulus, $G''$, of glycerol during heating and cooling temperature ramps, applied at a rate 20°C/min. The insets in (b) show the areas enclosed by the storage modulus *vs.* time curves during heating and cooling temperature ramps.

# References:

[1] Brookhaven Instruments Corporation, New York 11742, U.S.A., BI-200SM and BI-9000AT user manual, Brookhaven Instruments Corporation., 2nd Edition, available at https://www.brookhaveninstruments.com/wp-content/uploads/2023/08/BI-200SM-Manual-V.-2.0.pdf (Agust 1993).

[2] D. Saha, Y. M. Joshi, R. Bandyopadhyay, Investigation of the dynamical slowing down process in soft glassy colloidal suspensions: comparisons with supercooled liquids, Soft Matter 10 (18) (2014) 3292–3300.

[3] B. J. Berne, R. Pecora, Dynamic light scattering: with applications to chemistry, biology, and physics, Courier Corporation, 2000.

[4] A. Siegert, M. Rad, Lab report no. 465 (1943). 4. 5. 6. jh vanvleck and d. middleton, Proc. IEEE 54 (1966) 2.

[5] C. P. Lindsey, G. D. Patterson, Detailed comparison of the williams–watts and cole–davidson functions, The Journal of chemical physics 73 (7) (1980) 3348–3357.

[6] M. M. Cross, Rheology of non-newtonian fluids: a new flow equation for pseudoplastic systems, Journal of colloid science 20 (5) (1965) 417–437.

[7] S. K. Behera, D. Saha, P. Gadige, R. Bandyopadhyay, Effects of polydispersity on the glass transition dynamics of aqueous suspensions of soft spherical colloidal particles, Physical Review Materials 1 (5) (2017) 055603.

[8] C. Misra, S. K. Behera, R. Bandyopadhyay, Influence of particle size on the thermoresponsive and rheological properties of aqueous poly (n-isopropylacrylamide) colloidal suspensions, Bulletin of Materials Science 43 (1) (2020) 182.

[9] J. Mattsson, H. M. Wyss, A. Fernandez-Nieves, K. Miyazaki, Z. Hu, D. R. Reichman, D. A. Weitz, Soft colloids make strong glasses, Nature 462 (7269) (2009) 83–86.

[10] G. K. Batchelor, The effect of brownian motion on the bulk stress in a suspension of spherical particles, Journal of fluid mechanics 83 (1) (1977) 97–117.

[11] J. F. Brady, M. Vicic, Normal stresses in colloidal dispersions, Journal of Rheology 39 (3) (1995) 545–566.

[12] X. Di, X. Peng, G. B. McKenna, Dynamics of a thermo-responsive microgel colloid near to the glass transition, The Journal of chemical physics 140 (5) (2014).


    
\end{document}